\documentclass[aps,pre,onecolumn,floatfix,showkeys]{revtex4}

\usepackage{graphicx}
\usepackage{amssymb,amsfonts,amsmath}
\usepackage{epsf}
\usepackage{subfigure}
\usepackage{epstopdf}
\usepackage{mathrsfs}
\usepackage{color}

\usepackage{soul}
\usepackage{ulem}

\newcommand{\be}{\begin{equation}}
\newcommand{\ee}{\end{equation}}
\newcommand{\bea}{\begin{eqnarray}}
\newcommand{\eea}{\end{eqnarray}}

\begin{document}

\title{Thermodynamics and statistical mechanics of \\ chemically-powered synthetic nanomotors}

\author{Pierre Gaspard}
\email{gaspard@ulb.ac.be}
\affiliation{ Center for Nonlinear Phenomena and Complex Systems, Universit{\'e} Libre de Bruxelles (U.L.B.), Code Postal 231, Campus Plaine, B-1050 Brussels, Belgium}

\author{Raymond Kapral}
\email{rkapral@chem.utoronto.ca}
\affiliation{ Chemical Physics Theory Group, Department of Chemistry, University of Toronto, Toronto, Ontario M5S 3H6, Canada}

\begin{abstract}
Colloidal motors without moving parts can be propelled by self-diffusiophoresis, coupling molecular concentration gradients generated by surface chemical reactions to the velocity slip between solid Janus particles and the surrounding fluid solution.  The interfacial properties involved in this propulsion mechanism can be described by nonequilibrium thermodynamics and statistical mechanics, disclosing the fundamental role of microreversibility in the coupling between motion and reaction.  Among other phenomena, the approach predicts that propulsion by fuel consumption has the reciprocal effect of fuel synthesis by mechanical action.
\begin{figure}[h]
\centering
\resizebox*{5cm}{!}{\includegraphics{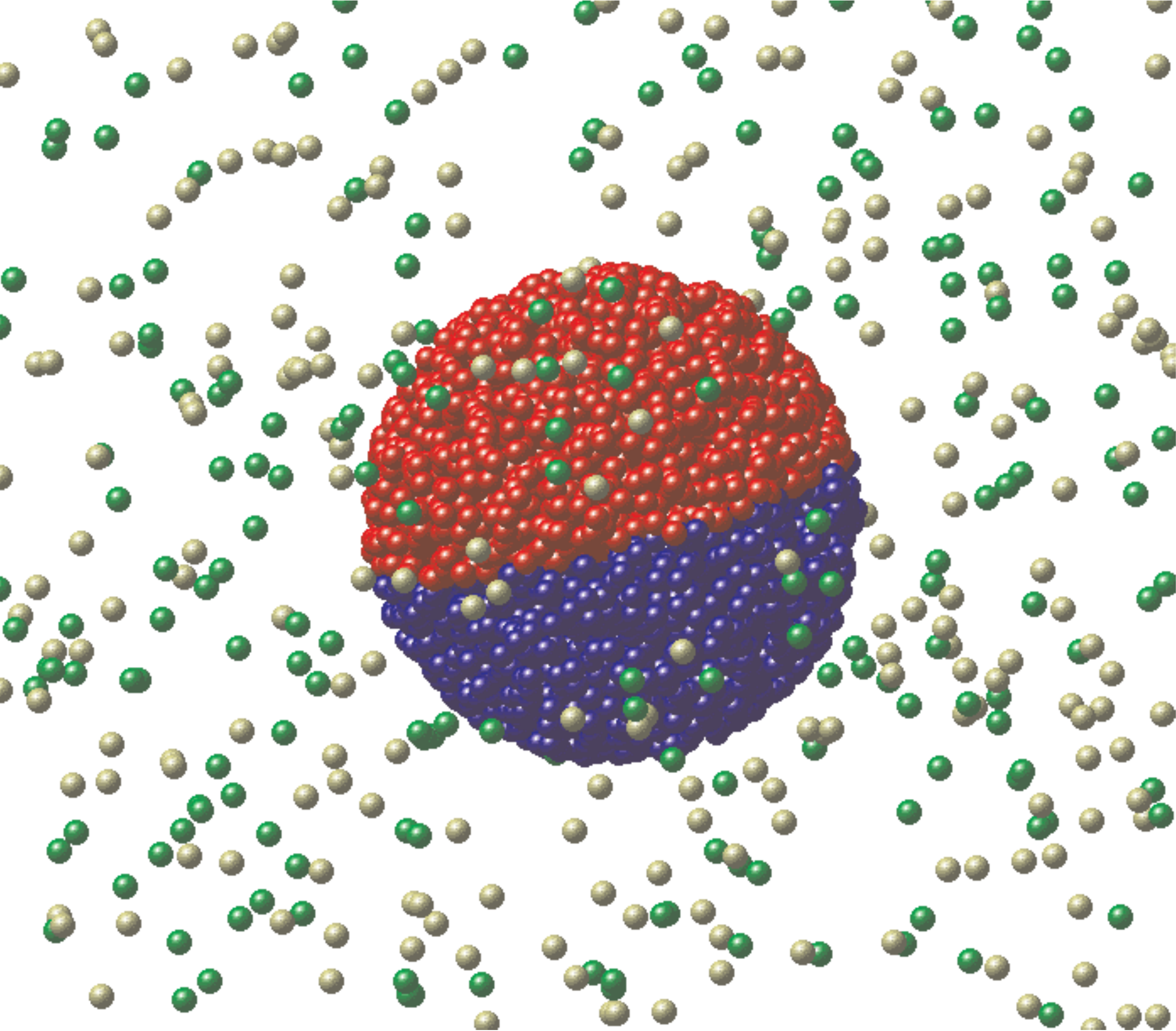}}
\end{figure}
\end{abstract}


\maketitle

\section{Introduction}

Artificial or natural engines, machines and motors with dimensions ranging from molecular to macroscopic scales are able to convert free energy into motion by various mechanisms that depend on their size.  At the smallest scale, we find synthetic molecular machines made of catenanes and rotaxanes.  These nanometric machines are driven by successive changes in the surrounding concentrations and/or by photochemical means \cite{SG11}.  On the same scale, organometallic complexes carrying out homogeneous catalysis \cite{vL04,DPBPHPS16,OWOG18} also undergo directional motion, in particular during polymerization, or enhanced diffusion.  These machines are so small that the substrate molecules remain in the periphery of their catalytic centers.

Biomolecular motors are larger molecular assemblies with enzymatic activity.  Examples are rotary motors such as the ATP synthase and bacterial flagellar rotary motors, and the actin-myosin, kinesin-microtubule, and dynein-microtubule linear motors \cite{ABJLRRW98,J04}.  Since they have sizes of the order of 10 nm or more, like other enzymes, they can accommodate internal catalytic sites or channels.  They are powered by ATP hydrolysis or transmembrane ionic currents.

Synthetic motors with sizes ranging from tens of nanometers to micrometers have been made and studied~\cite{W13}. They are solid particles composed of several materials and some portions of the surfaces of these motors are catalytically active.  The processes that drive propulsion take place at the interface between the solid particle and the surrounding fluid, and the properties of the interface thus play a key role in the mechanism. Since the thickness of the interface is much smaller than the particle diameter, a continuous-medium description may be used.  Indeed, it is known that continuous-medium equations such as the Navier-Stokes equations are valid for fluid flows down to the nanoscale \cite{MMPK88}.  At this level of description, the velocity or concentration fields should satisfy boundary conditions incorporating the relevant interfacial properties, which are themselves determined by the molecular structure and molecular dynamics of the interface.

The colloidal motors that are the focus of this review are paradigmatic examples of synthetic motors because their propulsion is autonomous once they are immersed in a suitable solution; in addition, the collective behavior of many motors has interesting properties. In this regard, a comparison with biological systems is inspiring.  In muscles, the actin-myosin linear motors compose sarcomeres, which are the intracellular structures performing work during muscle contraction.  Although the contraction velocity of a single myosin head running along an actin filament is of the order of a micrometer per second, the muscle can reach a velocity multiplied by the number of sarcomeres composing the muscular fibres.  This example shows what collective motion of micrometric machines can achieve by cooperative action.

The paper is organized as follows. The phenomenology of synthetic chemically-powered nanomotors is described in Section~\ref{sec:motors}.  Fluctuating chemohydrodynamics is developed in Section~\ref{chemohydro} and solved for spherical Janus particles in Section~\ref{sph-problem}, in order to obtain the coupled overdamped Langevin equations ruling their dynamics in Section~\ref{sec:Langevin}.  Fluctuation theorems and microreversibility are discussed in Section~\ref{sec:FT}.  A comparison with biomolecular motors is given in Section~\ref{sec:compare}.  Section~\ref{sec:collective} is devoted to the implications for collective behavior.  Concluding remarks are given in Section~\ref{sec:conclude}.

\section{Synthetic chemically-powered motors}
\label{sec:motors}

Some of the first micron-scale synthetic motors were bimetallic rods using hydrogen peroxide fuel \cite{PKOSACMLC04,FAMO05}. Subsequently motors with other shapes made from other materials were constructed and studied~\cite{W13,WDAMS13,SSK14,CRRK14,APTW14}. An important class of colloidal particles is made of silica spheres with a platinum cap. Heterogeneous catalysis will take place on this metallic surface once immersed in a solution containing some reactant, such as hydrogen peroxide.  These colloidal motors can be described by simplified models assuming the particle is spherical with a catalytic portion, or dimers composed of linked catalytic and non-catalytic spheres \cite{RK07}.

It is worth noting that liquid droplets can also exhibit self-propulsion \cite{SMHY05,CNY09,TSH11,IvLMD14}.  Here, the propulsion mechanism is the Marangoni effect due to the dependence of the surface tension between the droplet and the surrounding solution on tensio-active species.  The surface tension is an important interfacial property that determines the boundary conditions of the velocity fields inside and outside the droplet in the framework of hydrodynamics.  Although the surface tension as well as other interfacial properties are determined by the molecular structure of the interface, the interface is assumed to be arbitrarily thin in the continuous-medium description provided by hydrodynamics for the colloidal particles immersed in a fluid of interest here.

An important difference between machines with dimensions that are above and below micrometers is that the thermal and molecular fluctuations become negligible at macroscopic scales although their effects are essential on micrometric or submicrometric scales, as it is well known for Brownian motion.  Therefore, small machines and motors should be described in terms of the theory of stochastic processes and statistical mechanics.  Indeed these theories provide a way to describe random motions in terms of probabilities, mean values, variances, and other statistical moments or cumulants.   The effects of thermal fluctuations on continuous velocity or concentration fields can also be described, for instance, using fluctuating hydrodynamics \cite{LL80Part2,OS06}.  Instead, on macroscopic scales, the amplitudes of the thermal fluctuations become smaller than the size of the machine, so that their effects may be neglected.

A distinction should be made between motors without moving parts that are propelled by the reactions taking place at their interfaces, and those that are propelled by cyclic changes in their shapes~\cite{K13}. For motors such as bacteria propelled by flagellae, amoebae moving by the dynamics of their cytoskeleton, unicellular organisms swimming with ciliae, as well as animals and macroscopic engines, it is the time dependence of the shape that performs work and leads to propulsion. For motors without moving parts, propulsion has its origin in slippage of the velocity at the interface between the solid particle and the surrounding fluid due to phoretic effects that include thermophoresis involving temperature gradients along the interface, diffusiophoresis involving concentration gradients, and electrodiffusiophoresis in electrolyte solutions.  More specifically, these phoretic effects are induced by the molecular interaction forces between the solid particle and the solute species in the solution.  For these colloidal motors, surface reactions generate inhomogeneous concentration fields of the reactant and product species in the solution around the solid particle giving rise to  concentration gradients along the surface of the particle.  Through diffusiophoresis, these concentration gradients induce a slippage between the fluid and solid velocities, allowing the propulsion of the particle in the fluid.

A key issue is to understand the mechanism by which the transduction of the available free energy into motion takes place.  Every motor uses some source of free energy, which can be heat, chemical reaction, or electron transfer provided by the voltage of a battery. Biomolecular motors are often powered by the hydrolysis of adenosine triphosphate~(ATP) into adenosine diphosphate~(ADP) and an inorganic phosphate~(P$_{\rm i}$): ${\rm ATP}({\rm aq}) \to {\rm ADP}({\rm aq}) + {\rm P}_{\rm i}({\rm aq}), \;  \Delta G^0 = - 30.5 \, {\rm kJ/mol}$, taking place inside their catalytic sites and generating conformational changes of the motor protein complex. Colloidal motors made of metal catalyst often use hydrogen peroxide as fuel.  The free energy is provided by aqueous reaction, ${\rm H}_2{\rm O}_2({\rm aq}) \to {\rm H}_2{\rm O}({\rm l}) + \frac{1}{2} \, {\rm O}_2({\rm g}), \; \Delta G^0 = - 103.2 \, {\rm kJ/mol}$, taking place on their metal catalytic surfaces. With hydrogen peroxide fuel, micrometric colloidal particles can reach speeds of the order of 10~$\mu$m/s~\cite{VTZKGKO10,KYCS10}. Higher speeds of about 20~$\mu$m/s have been observed \cite{GPDW14} using hydrazine as fuel and iridium as catalyst.

These surface reactions, as well as diffusiophoresis, surface tension, or hydrophobicity are interfacial properties that can be tuned by changing the materials composing the interface.  Nonequilibrium thermodynamics provides a way to identify the material and interfacial properties that are involved in the mechanisms of propulsion.  The next section is devoted to this key issue.

\section{Fluctuating chemohydrodynamics}
\label{chemohydro}

\subsection{Catalytic particle moving in a reactive solution}

The goal is to find equations of motion for an active colloidal particle and its surrounding
fluid environment that account for thermal fluctuations as well as the processes that give rise to propulsion by diffusiophoresis. For inactive particles, this task reduces to the well-known problem of determining the stochastic evolution equations for Brownian motion. A similar approach can be adopted to account for active motion, and we shall illustrate it below by considering Janus motors whose surfaces have catalytic and non-catalytic hemispheres.

The starting point of the calculation is the computation of the force and torque exerted by the fluid on the Janus particle using methods developed in the seventies for Brownian motion~\cite{MB74,BM74}. The Navier-Stokes equations are linearized, which is justified since the flow is laminar around micrometric particles, and they are solved using the method of the induced force densities that efficiently takes the boundary conditions into account. We now outline how this method can be extended to include coupling to concentration fields so that it is able to describe diffusiophoretic effects (or thermophoretic effects from temperature gradients) on the motion of the Janus particle.

The force exerted on the Janus particle by the fluid is given by the surface integral of the pressure tensor ${\boldsymbol{\mathsf P}}$ at the interface $\Sigma(t)$ between the fluid and the Janus particle and, if present, an external force ${\bf F}_{\rm ext}$.  As a consequence, Newton's equation for the Janus particle takes the form,
\be
m\, \frac{d{\bf V}}{dt} = -\int_{\Sigma(t)} {\boldsymbol{\mathsf P}}({\bf r},t)\cdot{\bf n} \, d\Sigma + {\bf F}_{\rm ext},
\label{Langevin-Eq-0}
\ee
where $m=\int_{{\cal V}(t)}\rho_{\rm solid} \, d{\bf r}$ is the mass of the Janus particle and $\rho_{\rm solid}$ its mass density.  In a gravitational force field with acceleration ${\bf g}$, the external force is given by ${\bf F}_{\rm ext}=m{\bf g}$.

In a similar manner, a torque is exerted by the fluid on the Janus particle so that the angular velocity obeys the equation,
\be
{\boldsymbol{\mathsf I}}\cdot\frac{d\pmb{\Omega}}{dt} = -\int_{\Sigma(t)} \Delta{\bf r}\times \left[{\boldsymbol{\mathsf P}}({\bf r},t)\cdot{\bf n}\right]  d\Sigma + {\bf T}_{\rm ext},
\label{rot-Langevin-Eq-0}
\ee
where the inertia tensor ${\boldsymbol{\mathsf I}}$ of the Janus particle has the components $I_{ij}=\int_{{\cal V}(t)}\rho_{\rm solid} \big(\Delta{\bf r}^2\, \delta_{ij}-\Delta r_i\Delta r_j\big) d{\bf r}$ with $\Delta{\bf r}\equiv{\bf r}-{\bf R}(t)$, where ${\bf R}(t)$ is the position of the center of mass of the Janus particle, and ${\bf T}_{\rm ext}$ is an external torque \cite{H75,F76a,F76b,BAM77}.  The external torque may be due to an external magnetic field $\bf B$ exerted on a magnetic dipole $\mu{\bf u}$ attached to the particle \cite{BDLRVV16}, in which case ${\bf T}_{\rm ext} = \mu{\bf u}\times{\bf B}$, or due to a gravitational field acting on the nonuniform mass density of the Janus particle \cite{CE13}.

In order to compute the force and torque on the Janus particle we must evaluate the surface integrals involving the fluid pressure tensor at the catalytic surface. This computation requires a knowledge of the fluctuating fluid equations, along with boundary conditions that prescribe how the fluid fields are coupled to the concentration fields on the particle surface. These equations must be consistent with the basic conditions for microscopic time reversibility and other symmetries of the problem.  Nonequilibrium thermodynamics combined with the theory of stochastic processes provides a method that can be used to construct these equations.

\subsection{Thermodynamics and fluctuations}

In nonequilibrium thermodynamics, the densities, $a$, of any quantities such as the concentrations $c_k$ of particles of species $k$, the mass density $\rho$, the linear momentum density ${\bf g}=\rho{\bf v}$ where $\bf v$ is the fluid velocity, the energy density $\varepsilon=e+\rho{\bf v}^2/2$ including the internal energy density $e$ and the kinetic energy density $\rho{\bf v}^2/2$, as well as the entropy density $s$, obey balance equations with the general form,
\be
\partial_t a + \pmb{\nabla}\cdot\left( a {\bf v} + {\bf J}_a\right) = \sigma_a \, , \label{bal-eq-bulk}
\ee
where ${\bf J}_a$ are the corresponding current densities defined with respect to the barycentric motion and $\sigma_a$ are source densities \cite{P67,GM84,N79}.

Thermal fluctuations are incorporated in such continuum descriptions of the dynamics by adding noise terms $\delta J_\alpha(t)$ to the mean currents $\langle J_\alpha\rangle$ for the different irreversible processes $\{\alpha\}$ (the index $\alpha$ denoting the type and components of the scalar, vector or tensorial quantities): $J_\alpha = \langle J_\alpha\rangle + \delta J_\alpha(t)$~\cite{LL80Part1,LL80Part2,G04,OS06}. The mean currents are assumed to satisfy linear phenomenological laws,
\be
\langle J_\alpha\rangle = \sum_\beta L_{\alpha\beta}  A_\beta
\label{lin_laws}
\ee
in terms of linear response coefficients $L_{\alpha\beta}$ and the thermodynamic forces also called affinities $\{A_\alpha\}$ corresponding to the currents. The linear dependence~(\ref{lin_laws}) remains valid as long as the nonequilibrium driving processes take place over scales larger than the molecular mean free paths. According to microreversibility, the linear response coefficients obey the Onsager-Casimir reciprocal relations $L_{\alpha\beta}=\epsilon_\alpha\epsilon_\beta L_{\beta\alpha}$ where $\epsilon_\alpha=\pm1$ when $A_\alpha$ is even or odd under time reversal~\cite{O31a,O31b,C45,W67,BAM76,GM84,H69}.  Only coefficients $L_{\alpha\beta}$ that couple processes with the same parity under time reversal contribute to the entropy production rate,
\be
\frac{1}{k_{\rm B}}\frac{d_{\rm i}S}{dt} = \sum_\alpha \langle J_\alpha\rangle A_\alpha =
\sum_{\alpha,\beta} L_{\alpha\beta} A_\alpha A_\beta \geq 0 \, ,
\label{entrprod2}
\ee
where $k_{\rm B}$ is the Boltzmann constant. The fluctuating currents $\delta J_\alpha(t)$ are assumed to be Gaussian white noise processes characterized by
\be
\langle\delta J_\alpha(t)\rangle = 0, \quad
\langle\delta J_\alpha(t)\, \delta J_\beta(t')\rangle = (L_{\alpha\beta}+L_{\beta\alpha})\, \delta(t-t') \, ,
\label{Gaussian}
\ee
on time scales longer than molecular correlation times~\cite{LL80Part1,LL80Part2,G04,OS06}.  The second equations in (\ref{Gaussian}) are called fluctuation-dissipation relations~\cite{CW51}.  Notice that they vanish if processes with opposite parities under time reversal are coupled together by coefficients such that $L_{\alpha\beta}=-L_{\beta\alpha}$. In this case there is no associated noise to consider.

This general stochastic formulation may be applied to a Janus motor propelled by a diffusiophoretic mechanism. We suppose that the motor is suspended in a multi-component fluid containing solute species, labeled by the index $k$, that interact with the motor through short-range intermolecular potentials $u_k$.

We further assume that the reversible reactions ${\rm A} \rightleftharpoons{\rm B}$ occur on the catalytic hemisphere of the motor, and call species ${\rm A}$  the fuel and ${\rm B}$  the product. (Generalizations to other reaction schemes such as those mentioned in Sec.~\ref{sec:motors} can be carried out.) These chemical reactions produce inhomogeneous $c_{\rm A}$ and $c_{\rm B}$ concentration fields in the motor vicinity that lead to a body force on the motor. If no external forces act on the system and momentum is conserved, fluid flows arise in the surrounding medium and are responsible for motor propulsion. The forms that the stochastic equations for the fluid velocity and concentration fields take in the solution and on the surface are discussed below.

\vskip 0.2 cm

\subsection{Stochastic equations in the bulk phases}

The bulk fluid phase equations are well known. If the exothermicity of the reaction is negligible, we may suppose that the system remains isothermal with an invariant and uniform temperature $T$. The hydrodynamic and diffusive processes in the solution surrounding the catalytic Janus particle are described by the coupled Navier-Stokes and diffusion equations. The fluid is assumed to be incompressible, $\pmb{\nabla}\cdot{\bf v} = 0$, so that the mass density remains uniform.  The fluctuating Navier-Stokes equations for the velocity field $\bf v$ are given by
\be
\rho\left(\partial_t{\bf v} + {\bf v}\cdot\pmb{\nabla}{\bf v}\right) = -\pmb{\nabla}\cdot{\boldsymbol{\mathsf P}}=-\pmb{\nabla}\cdot(P\,{\boldsymbol{\mathsf 1}}+\pmb{\Pi}) \, ,
\label{NS_eqs}
\ee
where $\rho$ is the fluid mass density, $P$ is the hydrostatic pressure, and $\pmb{\Pi}$ is the viscous part of the pressure tensor, $\pmb{\Pi} = - \eta \left(\pmb{\nabla}{\bf v}+\pmb{\nabla}{\bf v}^{\rm T}\right)+\pmb{\pi}$,
where $\eta$ is the shear viscosity, which is related to the corresponding Onsager coefficient by $L/T=2\eta$. The Gaussian white noise fields $\pi_{ij}$ are characterized by $\langle \pi_{ij}({\bf r},t)\rangle = 0$ and $\langle \pi_{ij}({\bf r},t)\, \pi_{kl}({\bf r}',t')\rangle = 2 k_{\rm B}T \eta \left(\delta_{ik}\delta_{jl}+\delta_{il}\delta_{jk}\right)\delta({\bf r}-{\bf r}')\delta(t-t')$.

The fluctuating diffusion equations for the concentration fields $c_k$ of the different solute species $k=1,2,...$ have the form
\be
\partial_t \, c_k + {\bf v}\cdot\pmb{\nabla}c_k=-\pmb{\nabla}\cdot{\bf J}_k \, ,
\label{diff-eq-1}
\ee
where the current densities can be expressed as ${\bf J}_k =  - D_k \pmb{\nabla}c_k + \pmb{\eta}_k$,
in terms of the molecular diffusivity $D_k$ of species $k$ and Gaussian white noise fields $\pmb{\eta}_k$ satisfying $\langle \pmb{\eta}_k({\bf r},t)\rangle = 0$ and $\langle \pmb{\eta}_k({\bf r},t)\,   \pmb{\eta}_{k'}({\bf r}',t')\rangle =2 D_k c_k \delta_{kk'} \delta({\bf r}-{\bf r}') \, \delta(t-t')\, {\boldsymbol{\mathsf 1}}$, with ${\boldsymbol{\mathsf 1}}$ denoting the $3\times 3$ identity matrix \cite{G04,OS06}.  The correlation functions are written to account for the fact that the affinities associated with the diffusive part of current densities are given by the gradients of chemical potentials $\mu_k=\mu_k^0 +k_{\rm B}T \ln(c_k/c^0)$ so that The Onsager coefficients are given by $L_{kk'}/T=\delta_{kk'}D_k c_k/(k_{\rm B} T)$ \cite{GM84}. Moreover, the noise terms on the pressure and current densities are uncorrelated: $\langle \pi_{ij} ({\bf r},t)\,  \pmb{\eta}_{k'}({\bf r}',t')\rangle=0$.

Inside a solid Janus particle of radius $R$, the velocity and concentration fields take the values, ${\bf v}({\bf r},t) = {\bf V}(t) + \pmb{\Omega}(t)\times \left[ {\bf r}-{\bf R}(t)\right] \; \mbox{and} \; c_k({\bf r},t)=0$, for $\Vert{\bf r}-{\bf R}(t)\Vert <R$.

\subsection{Stochastic equations at the interface}

Any density $a$ can be decomposed as $a = a^+ \theta^+ + a^{\rm s} \delta^{\rm s} + a^- \theta^-$, in terms of the densities $a^{\pm}$ in the two bulk phases on both sides of the interface and the excess surface density $a^{\rm s}$. Here $\theta^\pm$ are Heaviside functions for the two bulk phases and the surface Dirac distribution $\delta^{\rm s}({\bf r},t)$ restricts the quantity it multiplies to the interface. Given this decomposition, stochastic equations that are consistent with microscopic reversibility can be written for interfacial quantities.

Several irreversible processes take place at the interface between the fluid and the Janus particle.  First, there is the reaction ${\rm A}\rightleftharpoons{\rm B}$ with stoichiometric coefficients $\nu_{\rm A}=-1$ and $\nu_{\rm B}=+1$.  This reaction takes place on the catalytic hemisphere of the Janus particle with local rate,
\be
w= \kappa_+ c_{\rm A}-\kappa_- c_{\rm B} + \xi^{\rm s} \, ,
\label{ws}
\ee
where $\kappa_\pm$ are the rate constants that are positive on the chemically active surface and zero elsewhere, and $\xi^{\rm s}({\bf r},t)$ is the interfacial noise associated with the surface reaction that satisfies $\langle\xi^{\rm s}({\bf r},t)\rangle = 0$ and $\langle\xi^{\rm s}({\bf r},t)\, \xi^{\rm s}({\bf r}',t')\rangle= \left(\kappa_+ c_{\rm A}+\kappa_- c_{\rm B}\right)\delta_{\bot}({\bf r}-{\bf r}') \, \delta(t-t')$, where ${\bf r}$ and ${\bf r}'$ are restricted to the interface by the delta distribution $\delta_{\bot}({\bf r}-{\bf r}')$.  We note that the rate constants $\kappa_{\pm}$ have the units of m/s.  Equivalently, we can use the rate constants $k_{\pm}^0\equiv 4\pi R^2\kappa_{\pm}$ along with the diffusive rate constants $k_{D_k}\equiv 4\pi R D_k$, which have the units of m$^3$/s.

Second, there is a coupling between the frictional force along the interface associated with the partial slip of the velocity field between the fluid and the solid particle, and the interfacial diffusive transport of the excess surface densities $c_k^{\rm s}=\Gamma_k$.  Diffusiophoresis results from the coupling between these two processes \cite{W67,BAM76,K77,B86}.  We have
\bea
{\bf n}\cdot{\pmb{\Pi}}\cdot{\boldsymbol{\mathsf 1}}_{\bot} &=& -\lambda\, {\bf v}_{\rm slip} - \sum_k \lambda b_k \,\pmb{\nabla}_{\bot}c_k + {\bf f}^{\rm s}_{\rm fl}\,  ,  \qquad \label{BC3+n}\\
{\bf J}_k^{\rm s} &=& \frac{\lambda b_k c_k}{k_{\rm B}T}\, {\bf v}_{\rm slip}  - D_k^{\rm s}\, \pmb{\nabla}_{\bot}\Gamma_k+ \pmb{\eta}^{\rm s}_{k}\, , \label{BC2+n}
\eea
where ${\bf n}$ is a unit vector normal to the interface, ${\boldsymbol{\mathsf 1}}_{\bot}={\boldsymbol{\mathsf 1}}-{\bf n}{\bf n}$, and the velocity slip between the fluid and the solid particle is given by ${\bf v}_{\rm slip} = {\boldsymbol{\mathsf 1}}_{\bot}\cdot\left\{{\bf v}({\bf r},t) - {\bf V}(t) - \pmb{\Omega}(t)\times \left[ {\bf r}-{\bf R}(t)\right] \right\}$, for $\Vert{\bf r}-{\bf R}(t)\Vert =R$. Equation~(\ref{BC3+n}) provides a boundary condition on the ${\boldsymbol{\mathsf 1}}_{\bot}\cdot{\bf v}$ component of the velocity field at the interface with the Janus particle. In addition to this boundary condition, the normal component of the velocity field obeys ${\bf n}\cdot{\bf v}({\bf r},t) = {\bf n}\cdot{\bf V}(t)$ also at the interface \cite{ABM75}.

Equations~(\ref{BC3+n}) and (\ref{BC2+n}) are simplifications of more general results~\cite{GK18b} in which we assumed that the boundary layer and neighboring bulk phase are in local equilibrium so that the surface chemical potentials take their bulk values, $\mu_k^{\rm s}=\mu_k=\mu_k^0 +k_{\rm B}T \ln(c_k/c^0)$. In this case the surface Onsager coefficients are related to the coefficient of sliding friction, $L^{\rm s}_{\rm vv}/T=\lambda$, and the surface diffusion coefficients, $L^{\rm s}_{kl}/T=\delta_{kl} D^{\rm s}_k \Gamma_k/(k_{\rm B}T)$, while the surface Onsager coefficients relating the slip velocity and the surface current density are given by
$L^{\rm s}_{{\rm v}k}/T=- L^{\rm s}_{k{\rm v}}/T =  \lambda \, b_k \, c_k /(k_{\rm B} T)$
where the $b_k$ are diffusiophoretic constants and are given by
\be
b_k = \frac{k_{\rm B}T}{\eta} \left( K_k^{(1)} + b\, K_k^{(0)}\right) .
\label{b_k}
\ee
Here the slip length is $b=\eta/\lambda$ and $K_k^{(n)} \equiv \int_0^{\delta} dz \, z^n \, \left[ {\rm e}^{-\beta u_k(z)}-1 \right]$, where $\delta$ is the interfacial thickness \cite{A89,AP91,AB06}.

The Gaussian white noise terms satisfy the conditions $\langle{\bf f}_{\rm fl}^{\rm s}({\bf r},t)\rangle = 0$, $\langle{\bf f}_{\rm fl}^{\rm s}({\bf r},t)\, {\bf f}_{\rm fl}^{\rm s}({\bf r}',t')\rangle = 2 k_{\rm B}L^{\rm s}_{\rm vv}\, \delta_{\bot}({\bf r}-{\bf r}') \, \delta(t-t') \, {\boldsymbol{\mathsf 1}}_{\bot}$ and $\langle{\pmb{\eta}}_k^{\rm s}({\bf r},t)\rangle = 0$, $\langle{\pmb{\eta}}_k^{\rm s}({\bf r},t)\, {\pmb{\eta}}_l^{\rm s}({\bf r}',t')\rangle= 2 k_{\rm B}L^{\rm s}_{kl}\, \delta_{\bot}({\bf r}-{\bf r}') \, \delta(t-t') \, {\boldsymbol{\mathsf 1}}_{\bot}$, where ${\bf r}$ and ${\bf r}'$ are here also restricted to the interface. Since the interfacial slippage has an affinity that is odd under time reversal while the gradients of interfacial chemical potentials are even, the Onsager-Casimir reciprocal relations imply that the diffusiophoretic coupling terms do not appear in the expression for the excess surface entropy production, and the corresponding noise terms are uncorrelated, $\langle{\bf f}_{\rm fl}^{\rm s}({\bf r},t)\, {\pmb{\eta}}_k^{\rm s}({\bf r}',t')\rangle= 0$. (We may make an alternative choice taking $-{\bf v}_{\rm slip}/T$ as the current and ${\bf n}\cdot{\pmb{\Pi}}\cdot{\boldsymbol{\mathsf 1}}_{\bot}$ that is even under time reversal as the associated affinity.  With this choice the linear response coupling coefficients obey the standard symmetric Onsager reciprocal relations, and the diffusiophoretic coupling terms do appear in the expression for the excess surface entropy production.)

The boundary conditions on the concentration fields are determined by the transport of species in the direction normal to the interface, the surface reaction rate~(\ref{ws}), as well as the reciprocal effect of diffusiophoresis. For species $k$, the boundary condition can be expressed as
\be
{\bf n}\cdot{\bf J}_k =\nu_{k} w   - \Sigma^{\rm s}_k \, ,
\label{BC1bis}
\ee
also for $\Vert{\bf r}-{\bf R}(t)\Vert =R$. Here $\Sigma^{\rm s}_k =\partial_t\Gamma_k +\pmb{\nabla}_{\bot}\cdot(\Gamma_k{\bf v}^{\rm s}+{\bf J}_k^{\rm s})$ is a sink into the boundary pool of adsorbed species with ${\bf v}^{\rm s}$ the surface velocity, and ${\bf J}_k^{\rm s}$ the surface current density~(\ref{BC2+n}), which is determined, in particular, by the reciprocal effect of diffusiophoresis.

\subsection{Langevin equations}

Using the above results for the fluctuating fluid fields along with the boundary conditions for these fields on the surface of the particle, computation of the surface averages yields generalized Langevin equations for the force and torque.

For simplicity, we consider here time scales that are longer than the characteristic time of sound, $t_{\rm sound} \approx R/v_{\rm sound} \sim10^{-9}\, {\rm s}$, so that the fluid is effectively incompressible.  In addition, on time scales longer than the hydrodynamic time, $t_{\rm hydro} \approx R^2\rho_{\rm fluid}/\eta \sim 10^{-6}\, {\rm s}$, friction may be supposed to be time independent and given by the Stokes friction coefficient. This is the case for micrometric particles. Under these conditions, the Langevin equations take the form~\cite{GK18a},
\be
m\frac{d{\bf V}}{dt} = -\gamma_{\rm t}\, {\bf V}  + {\bf F}_{\rm d} + {\bf F}_{\rm ext} + {\bf F}_{\rm fl}(t) \, ,
\label{Langevin-eq-transl}
\ee
where $\gamma_{\rm t}= 6\pi\eta R (1+2b/R)/(1+3b/R)$ is the translational friction coefficient for arbitrary slip, and the diffusiophoretic force is
\be
{\bf F}_{\rm d}(t) = \frac{6\pi\eta R}{1+3b/R} \sum_k  \overline{b_k \,{\boldsymbol{\mathsf 1}}_{\bot}\cdot\pmb{\nabla}c_k({\bf r},t)}^{\rm s} ,
\label{Fd-surf-av}
\ee
expressed in terms of the surface average $\overline{(\cdot)}^{\rm s} = (4\pi R^2)^{-1} \int_{r=R} (\cdot ) \, d\Sigma$.
In this equation, we kept the deterministic diffusiophoretic force, so that the fluctuating force ${\bf F}_{\rm fl}(t)$ also includes any possible noisy contributions from the concentration field in addition to those from fluid fluctuations.

Similarly, from the computation of the torque exerted by the fluid on the particle we find the Langevin equation for the angular velocity,
\be
{\boldsymbol{\mathsf I}}\cdot\frac{d\pmb{\Omega}}{dt} = - \gamma_{\rm r}\,\pmb{\Omega} + {\bf T}_{\rm d} + {\bf T}_{\rm ext} + {\bf T}_{\rm fl}(t) \, ,
\label{Langevin-eq-rot}
\ee
where $\gamma_{\rm r}=8\pi\eta  R^3/(1+3b/R)$ is the time-independent rotational friction coefficient,
\be \label{rot-diff-t}
{\bf T}_{{\rm d}}(t)= \frac{12\pi\eta  R}{1+3b/R} \, \sum_k  \overline{b_k \, {\bf r}\times\pmb{\nabla}c_k({\bf r},t)}^{\rm s} ,
\ee
is the deterministic diffusiophoretic torque and ${\bf T}_{\rm fl}(t)$ is a random torque. The diffusiophoretic constants $b_k$ are included in the surface averages since they may be non-uniform on the particle surface in general.

\section{Spherical Janus particle}
\label{sph-problem}

\subsection{Diffusion and reaction} \label{sph-problem-RD}

For a micron-size Janus particle moving at the velocity $V_{\rm sd}\sim 10^{-5}\, {\rm m/s}$ in an aqueous solution where the solute molecular diffusion coefficients are of order $D_k\sim 10^{-9}\, {\rm m}^2/{\rm s}$, the P\'eclet numbers,
${\rm Pe}_k\equiv V_{\rm sd}R/D_k \sim 10^{-2}$, take values smaller than unity.  In this regime, the diffusion equation~(\ref{diff-eq-1}) reduces to $\nabla^2c_k=0$ because the term $\partial_t c_k+{\bf v}\cdot\pmb{\nabla}c_k$ is relatively proportional to the P\'eclet number while the noise term goes as its square root.  Discarding the noise and sink term $\Sigma^{\rm s}_k$, the boundary conditions~(\ref{BC1bis}) on the concentration fields become
\be\label{conc-bc}
D_k \partial_r c_k\vert_{r=R} = -\nu_k\, \chi(\theta) \, \left(\kappa_+ c_{\rm A}-\kappa_- c_{\rm B}\right)_{r=R}, \quad c_k\vert_{r=\infty} = \bar{c}_k \, ,
\ee
with $\chi(\theta) = H(\cos\theta)$ where the Heaviside function $H(\xi)$ takes the values $H(\xi)=1$ on the catalytic hemisphere and $H(\xi)=0$ on the chemically inactive hemisphere, and $\bar{c}_k$ are the uniform concentrations in the solution far from the particle.  We see that there is no coupling to the velocity field, so that the concentration fields can be determined at this level of approximation independently of the velocity field.  This approximation is justified because diffusion is fast enough around a micrometric particle for the concentration profiles to adjust themselves with respect to the orientation of the particle.  Indeed, the time scales for rotational or translational diffusion, $t_{\rm rot}^{\rm (D)} \approx 1/(2D_{\rm r}) \sim 3\, {\rm s}$ and $t_{\rm trans}^{\rm (D)} \approx R^2/D_{\rm t}\sim 5\, {\rm s}$, respectively, are longer than that for molecular diffusion, $t_{\rm mol}^{\rm (D)} \approx R^2/D_k \sim 10^{-3}\, {\rm s}$.

The concentration fields of species $k$ around a spherical Janus particle of radius $R$ are given by
\be \label{cA-f}
c_k(r,\theta) =  \bar{c}_k +\nu_k\frac{R}{D_k} \Big(\kappa_+\bar{c}_{\rm A}-\kappa_-\bar{c}_{\rm B}\Big) f(r,\theta) \, ,
\ee
where the function $f(r,\theta)$ satisfies the diffusion equation $\nabla^2f=0$, which can be solved by expressing it in a series of Legendre polynomials, $f(r,\theta) =\sum_{l=0}^{\infty} a_l \, P_l(\xi) (R/r)^{l+1}$,  with the boundary conditions $R\,\partial_rf\vert_R = H(\cos\theta)\left( {\rm Da}\, f-1\right)_R$ and $f\vert_{\infty}=0$ in spherical coordinates $(r,\theta,\phi)$ aligned parallel to the particle axis.  The boundary condition at the particle surface $r=R$ involves the dimensionless Damk\"ohler number, ${\rm Da} \equiv R (\kappa_+/D_{\rm A} + \kappa_-/D_{\rm B})$.  In the reaction-limited regime $k_\pm^0 \ll k_{D_k}$ and ${\rm Da}\ll 1$, while in the diffusion-controlled regime $k_\pm^0 \gg k_{D_k}$ and ${\rm Da}\gg 1$.

\begin{figure}[h]
\centering
\resizebox*{8cm}{!}{\includegraphics{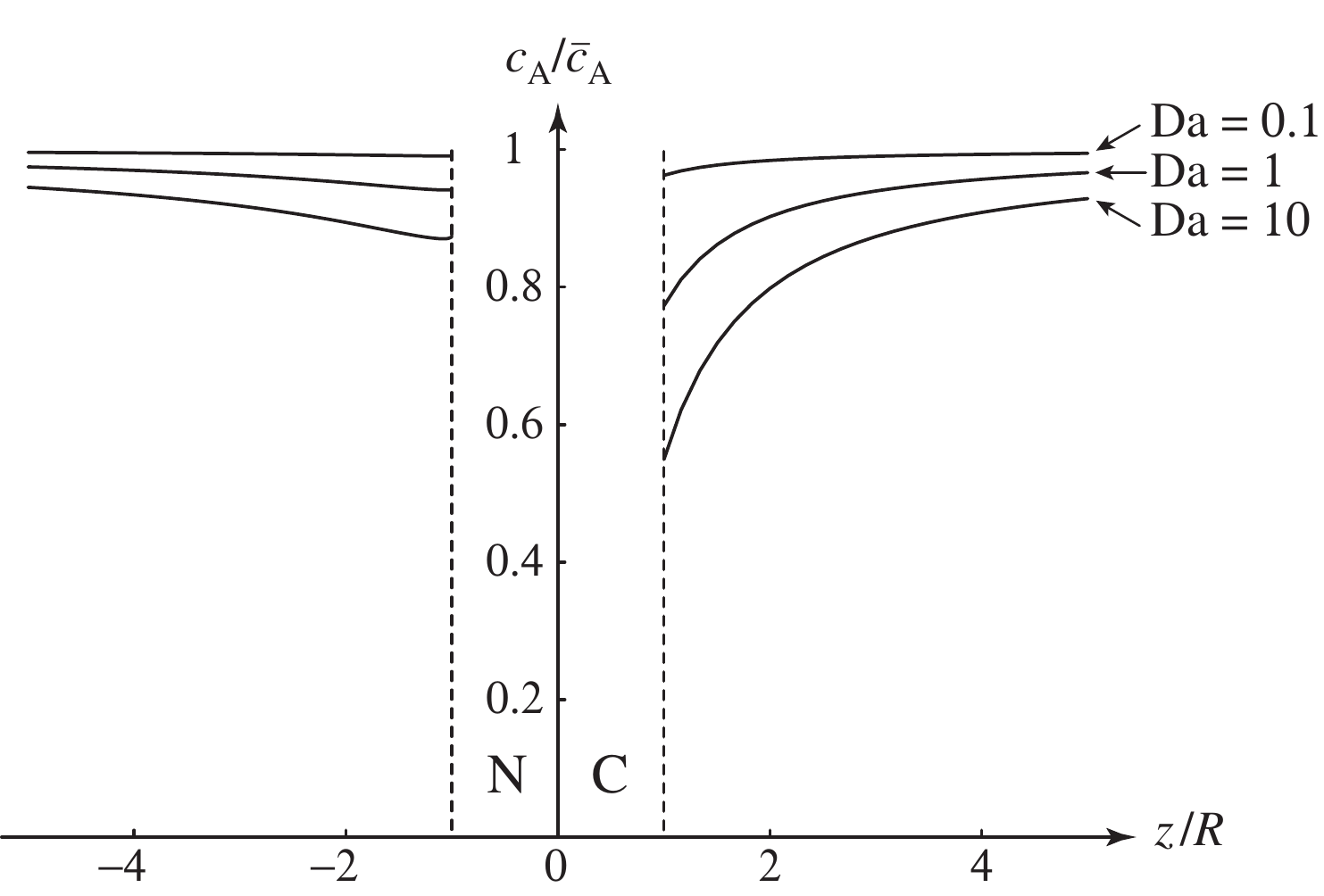}}
\caption{Relative concentration profiles of the fuel A along the axis $z$ of the Janus particle for three values of the Damk\"ohler number.  C and N denote the catalytic and non-catalytic hemispheres, respectively.  The profiles are obtained for $\bar{c}_{\rm B}=0$.} \label{fig1}
\end{figure}

Examples of concentration profiles for the fuel A are shown in Fig.~\ref{fig1} along the axis $z$ of the Janus particle.  We see the depletion of fuel near the catalytic hemisphere as the Damk\"ohler number increases from the reaction- to the diffusion-limited regime.  In this latter regime, we have that $f(r,\theta)\simeq({\rm Da}+1)^{-1}(R/r)$ near the catalytic hemisphere.  Accordingly, the relative fuel concentration behaves as $(c_{\rm A}/\bar{c}_{\rm A})_{\rm cat}\simeq ({\rm Da}+2)/(2{\rm Da}+2)$ and reaches the value $(c_{\rm A}/\bar{c}_{\rm A})_{\rm cat}\simeq 1/2$ in the limit ${\rm Da}\to\infty$.  In the same limit, the relative product concentration $(c_{\rm B}/\bar{c}_{\rm A})_{\rm cat}\simeq {\rm Da}/(2{\rm Da}+2)$ also reaches the value $(c_{\rm B}/\bar{c}_{\rm A})_{\rm cat}\simeq 1/2$.  We note that these concentration fields are obtained by averaging over the fluctuations.

Using these results, the mean value of the reaction rate, $W= \int_{\Sigma(t)} w \, d\Sigma$, where the surface integral is carried out over the catalytic hemisphere of the Janus particle because the rate constants $\kappa_{\pm}$ vanish on the non-catalytic hemisphere, may be written as
\be
W_{\rm rxn} = \Gamma \left(\kappa_+\bar{c}_{\rm A} -  \kappa_-\bar{c}_{\rm B}\right) ,
\label{Wrxn}
\ee
with $\Gamma=4\pi R^2 a_0$ where $a_0$ is the first coefficient of the expansion of $f(r,\theta)$, which behaves approximately as $a_0\simeq 0.5/(1+0.708\, {\rm Da})$ with the Damk\"ohler number.

At thermodynamic equilibrium, the concentrations $\bar{c}_{k}$ should satisfy the Guldberg-Waage condition, $\bar{c}_{\rm A, eq}/\bar{c}_{\rm B, eq} = \kappa_-/\kappa_+ = \exp[\Delta\mu^0/(k_{\rm B}T)]$,
where $\Delta\mu^0=\mu^0_{\rm B}-\mu^0_{\rm A}$ is the standard free energy of the reaction ${\rm A}\to{\rm B}$.  The free energy of the reaction is related to the concentrations by $\Delta\mu=\Delta\mu^0 + k_{\rm B}T \, \ln(\bar{c}_{\rm B}/\bar{c}_{\rm A})$.

The reaction is driven out of equilibrium if the concentrations are not in their equilibrium ratio.  In this respect, the nonequilibrium control parameter of the reaction is defined in general as the dimensionless chemical affinity
\be
A_{\rm rxn} \equiv \ln \frac{\kappa_+\bar{c}_{\rm A}}{\kappa_-\bar{c}_{\rm B}} = -\frac{\Delta\mu}{k_{\rm B}T} \, ,
\ee
which is positive (resp. negative) for the reaction running in the direction ${\rm A}\to{\rm B}$ (resp. ${\rm B}\to{\rm A}$), and vanishes at equilibrium since $\Delta\mu_{\rm eq}=0$.

In the following, we consider the reaction in the linear regime close to equilibrium where the deviations of the concentrations from their equilibrium values, $\delta \bar{c}_k \equiv \bar{c}_k-\bar{c}_{k,{\rm eq}}$ are small: $\vert\delta \bar{c}_k\vert \ll \bar{c}_{k,{\rm eq}}$.  In this regime, the chemical affinity can be approximated as $A_{\rm rxn} \simeq (\delta\bar{c}_{\rm A}/\bar{c}_{\rm A, eq}) - (\delta\bar{c}_{\rm B}/\bar{c}_{\rm B, eq})$, up to terms of second order in the concentration deviations $\delta \bar{c}_k$. Introducing the reaction diffusivity $D_{\rm rxn} \equiv \Gamma\left(\kappa_+\bar{c}_{\rm A} + \kappa_-\bar{c}_{\rm B}\right)/2$ associated with the reaction rate~(\ref{Wrxn}), the chemical affinity may also be written close to equilibrium as $A_{\rm rxn} = W_{\rm rxn}/D_{\rm rxn}$, up to terms with higher powers in the reaction rate.

\subsection{Fluid flow}

Once the concentration fields are known, the velocity field can be calculated to get the force and the torque on the particle. For micrometric particles moving at self-diffusiophoretic velocity $V_{\rm sd}\simeq 10\, \mu$m/s in an aqueous solution of viscosity $\eta\simeq 10^{-3}\, {\rm N\, s}/{\rm m}^2$, the fluid flow is laminar.  Indeed, the Reynolds number ${\rm Re}\equiv V_{\rm sd}R/\nu \sim 10^{-5}$ with the kinematic viscosity $\nu\simeq 10^{-6}\, {\rm m}^2/{\rm s}$ is much smaller than unity.  In this regime, the Navier-Stokes equations~(\ref{NS_eqs}) reduce to the Stokes equations $\eta \nabla^2{\bf v}=\pmb{\nabla}P$ because the left-hand side of~(\ref{NS_eqs}) is relatively proportional to the Reynolds number, while the noise term behaves as the square root of the Reynolds number.

\begin{figure}[htbp]
\centering
\resizebox*{5.5cm}{!}{\includegraphics{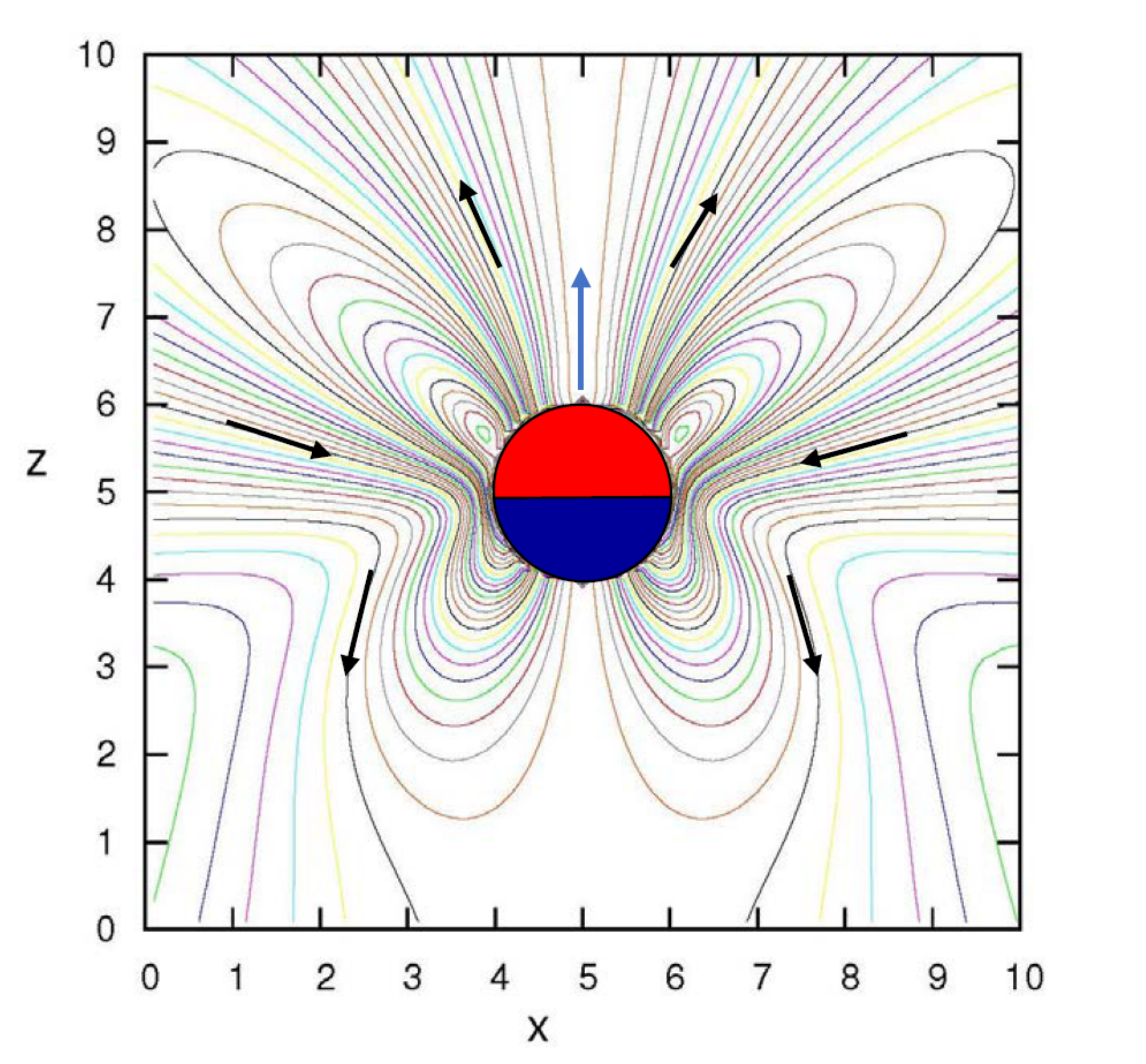}}
\caption{Streamlines of the velocity field~(\ref{flow_Janus}) around a Janus particle in the laboratory frame where the particle moves with the velocity $V_{\rm sd} = 2\Upsilon a_1/3$ for ${\rm Re}\simeq 0.013$, ${\rm Pe}\simeq 0.17$, and ${\rm Da}\simeq 6.2$ from~\cite{RHSK16}.  The vertical axis $z$ corresponds with ${\bf u}$ of the Janus particle.  The upper hemisphere is catalytic.  The motor here behaves as a pusher with $ \Upsilon a_2<0$. The blue arrow denotes the direction of particle motion and black arrows the fluid flow.} \label{fig2}
\end{figure}

Using the methods of~\cite{L52,B71}, for stick boundary conditions ($b=0$), and uniform diffusiophoretic constants~$b_{\rm A}$ and~$b_{\rm B}$, the velocity field is given by the solution of Stokes equations as~\cite{RHSK16,CEIG18}
\bea
&&{\bf v} = \Upsilon\, a_1 \left(\frac{R}{r}\right)^3 \left( {\bf n \, n}\cdot{\bf u} -\frac{1}{3}\, {\bf u}\right) + \frac{\Upsilon}{2} \sum_{l=2}^{\infty} l(l+1) \, a_l \left[\left(\frac{R}{r}\right)^{l+2}-\left(\frac{R}{r}\right)^l\right] P_l({\bf n}\cdot{\bf u}) \, {\bf n} \nonumber\\
&& -\frac{\Upsilon}{2}  \sum_{l=2}^{\infty} a_l \left[l \left(\frac{R}{r}\right)^{l+2}-(l-2)\left(\frac{R}{r}\right)^l\right] P_l^1({\bf n}\cdot{\bf u}) \left( {\bf n \, n}\cdot{\bf u} - {\bf u}\right) ,
\label{flow_Janus}
\eea
where ${\bf n}={\bf r}/r$, $\bf u$ is the unit vector in the direction of the Janus particle axis, $P_l^m(\xi)$ are the associated Legendre polynomials of degree $l$ and order $m$, $\Upsilon \equiv (b_{\rm B}/D_{\rm B} - b_{\rm A}/D_{\rm A}) (\kappa_+ \bar{c}_{\rm A}-\kappa_- \bar{c}_{\rm B})$, and $a_l$ are again the coefficients of the expansion of $f(r,\theta)$.
An example of the velocity field~(\ref{flow_Janus}) is depicted in Fig.~\ref{fig2} \cite{RHSK16}.

At large distances, the velocity field~(\ref{flow_Janus}) behaves as the Stokes dipole
\be
\mbox{self-diffusiophoresis driving:} \quad {\bf v} = -\frac{3}{2}\, \Upsilon \, a_2 \left(\frac{R}{r}\right)^2 \left[ 3\, ({\bf n}\cdot{\bf u})^2-1\right] {\bf n} + O(r^{-3}) \, ,
\label{v_far_field}
\ee
which is a puller if $ \Upsilon a_2>0$ and a pusher if $ \Upsilon a_2<0$.  We notice that this velocity field goes as $r^{-2}$ and thus contributes to the pressure tensor.  In contrast, if there are concentration gradients ${\bf g}_k$ but no reaction, the velocity field is exactly given by
\be
\mbox{diffusiophoresis driving:} \quad {\bf v} = \frac{1}{2} \left(\frac{R}{r}\right)^3 \left( 3\, {\bf n\, n }\cdot{\bf V}_{\rm d}-{\bf V}_{\rm d}\right)  \, ,\hspace{3cm}
\label{v-diff}
\ee
where ${\bf V}_{\rm d} = b_{\rm A}\, {\bf g}_{\rm A} + b_{\rm B}\, {\bf g}_{\rm B}$ is the diffusiophoretic velocity~\cite{A86,A89}.  In this passive case, the velocity field~(\ref{v-diff}) decreases as $r^{-3}$ faster than in the active case~(\ref{v_far_field}).

We recall that, if an external force ${\bf F}_{\rm ext}$ is exerted on the colloidal particle, the velocity field is given by
\be
\mbox{external force driving:} \quad {\bf v} = \frac{3}{4} \frac{R}{r} \left( {\bf V}+ {\bf n\, n }\cdot{\bf V}\right) +\frac{1}{4} \left(\frac{R}{r}\right)^3 \left( {\bf V}-3\,  {\bf n\, n }\cdot{\bf V}\right) ,
\ee
where ${\bf V}={\bf F}_{\rm ext}/\gamma_{\rm t}$ is the particle velocity.  Here, the velocity field decreases as $r^{-1}$.  The behavior as $r^{-2}$ for the particle driven by active self-diffusiophoresis is thus intermediate between its driving by an external force and by passive diffusiophoresis.  We note that ${\bf v}\vert_{r=R}={\bf V}$ in all the cases.

\subsection{Force and torque}

Since the concentration fields generated by the reaction are given by Eq.~(\ref{cA-f}), we may write an explicit formula for the deterministic part of the force~(\ref{Fd-surf-av}), giving the following self-diffusiophoretic force for arbitrary slip if the diffusiophoretic constants $b_k$ are uniform on the spherical surface of the Janus particle:
\be \label{diff-force}
{\bf F}_{\rm sd} = \frac{4\pi\eta R}{1+3b/R} \left(\frac{ b_{\rm B}}{D_{\rm B}}-\frac{ b_{\rm A}}{D_{\rm A}}\right)\left(\kappa_+\bar{c}_{\rm A} -  \kappa_-\bar{c}_{\rm B}\right)  a_1 {\bf u} \, .
\ee
For the same particle the diffusiophoretic torque~(\ref{rot-diff-t}) vanishes by cylindrical symmetry, so that ${\bf T}_{\rm sd}=0$ and only the frictional torque due to viscosity remains.

\section{Coupled overdamped Langevin equations}
\label{sec:Langevin}

For micrometer-sized particles inertia is unimportant and henceforth we consider the overdamped limits of the Langevin equations. Letting ${\bf V}=d{\bf r}/dt$, the Langevin equation for a spherical Janus particle reduces to
\be
\frac{d{\bf r}}{dt} = \beta D_{\rm t} \, {\bf F}_{\rm ext} + {\bf V}_{\rm sd}  + {\bf V}_{\rm fl}(t) \, ,
\label{eq-r}
\ee
in terms of the self-diffusiophoretic velocity ${\bf V}_{\rm sd} ={\bf F}_{\rm sd}/\gamma_{\rm t}$, the translational diffusion coefficient of the particle $D_{\rm t}=k_{\rm B} T/\gamma_{\rm t}$,
and the fluctuating velocity ${\bf V}_{\rm fl}(t) ={\bf F}_{\rm fl}(t)/\gamma_{\rm t}$ satisfying $\langle {\bf V}_{\rm fl}(t)\rangle = 0$ and $\langle {\bf V}_{\rm fl}(t)\, {\bf V}_{\rm fl}(t')\rangle = 2D_{\rm t} \, \delta(t-t') \, {\boldsymbol{\mathsf 1}}$.  The self-diffusiophoretic velocity can be written as ${\bf V}_{\rm sd}=V_{\rm sd} {\bf u}$ in terms of the unit vector $\bf u$,  and for a spherical Janus particle $V_{\rm sd}$ is given by
\be \label{eq:Vsd}
V_{\rm sd} = \frac{2}{3(1+2b/R)} \Big(\frac{b_{\rm B}}{D_{\rm B}}-\frac{b_{\rm A}}{D_{\rm A}}\Big)(\kappa_+ \bar{c}_{\rm A}-\kappa_- \bar{c}_{\rm B})a_1 \equiv \chi W_{\rm rxn}.
\ee
In the last equality we expressed the velocity in terms of the mean reaction rate~(\ref{Wrxn}) and the diffusiophoretic parameter, $\chi \equiv  F_{\rm sd}/(\gamma_{\rm t} W_{\rm rxn})$. The self-diffusiophoretic velocity~(\ref{eq:Vsd}) and the parameter $\chi$ remain finite in the limits of perfect stick ($b\to 0$) and perfect slip ($b\to\infty$) boundary conditions.

The overdamped limit of the rotational Langevin equation is
\be
\frac{d{\bf u}}{dt} = -\frac{1}{\gamma_{\rm r}} \, {\bf u}\times \left[{\bf T}_{\rm ext} + {\bf T}_{\rm fl}(t) \right]  \, .
\label{eq-rot}
\ee
The rotational diffusion coefficient is related to the rotational friction coefficient by $D_{\rm r}\equiv k_{\rm B}T/\gamma_{\rm r}$. Since the particle is spherical, this equation does not depend on the particle position or the reactive state (if the external torque is spatially uniform).  Consequently, this stochastic equation is autonomous; thus, it drives the direction $\bf u$ independently of what happens for translation and reaction.

In addition to these two equations we must also consider the stochastic equation for reaction since the concentration fields depend on reactive processes on the surface of the particle. The number $n$ of reactive events during the time interval $[0,t]$ since the beginning of observation satisfies the stochastic differential equation:
\be
\frac{dn}{dt} = W_{\rm sd} + W_{\rm rxn} + W_{\rm fl}(t) \, ,
\label{eq-n}
\ee
where, in addition to the mean rate $W_{\rm rxn}$, we have a contribution $W_{\rm sd}$ from self-diffusiophoresis to be determined. The fluctuating rate $W_{\rm fl}(t)$ satisfies $\langle W_{\rm fl}(t)\rangle = 0$ and $\langle W_{\rm fl}(t)\, W_{\rm fl}(t')\rangle = 2D_{\rm rxn} \, \delta(t-t')$.  The rate~(\ref{eq-n}) can be written as $dn/dt= -dN_{\rm A}/dt= dN_{\rm B}/dt$ in terms of the numbers of molecules A and B in the solution.

Equations~(\ref{eq-r}) and (\ref{eq-n}) are coupled equations for the currents $d{\bf X}/dt=({\bf J}_{\bf r}, J_n)$ of the variables ${\bf X}=({\bf r},n)$ that are associated with the corresponding mechanical and chemical affinities ${\bf A} =({\bf A}_{\rm mech} = \beta \, {\bf F}_{\rm ext},A_{\rm rxn} = W_{\rm rxn}/D_{\rm rxn})$. We can use the Onsager symmetry principle in order to determine $W_{\rm sd}$. Since the variables $\bf r$ and $n$ are even under time reversal, the Onsager coefficients must satisfy $L_{\alpha\beta}=L_{\beta \alpha}$. In view of this property the coupled equations must take the form,
\be \label{CL}
\frac{d{\bf X}}{dt}=\left(
\begin{array}{cc}
D_{\rm t} \, {\boldsymbol{\mathsf 1}} & \chi\, D_{\rm rxn} \, {\bf u} \\
\chi\, D_{\rm rxn} \, {\bf u} & D_{\rm rxn}
\end{array}
\right)\cdot{\bf A}+ \delta{\bf J}(t) ={\boldsymbol{\mathsf L}}\cdot{\bf A}+\delta{\bf J}(t)
\ee
where the vector of the noise terms is denoted by $\delta{\bf J}(t)$. Onsager symmetry dictates that $W_{\rm d}=\chi D_{\rm rxn} {\bf u}\cdot{\bf A}_{\rm mech}=\beta\chi D_{\rm rxn} {\bf u}\cdot{\bf F}_{\rm ext}$. From the fluctuation-dissipation relations~(\ref{Gaussian}), we also have the property, $\langle {\bf V}_{\rm fl}(t)\, W_{\rm fl}(t')\rangle = 2\chi D_{\rm rxn} \, {\bf u} \, \delta(t-t')$, for the noise correlations.

\begin{figure}
\centering
\resizebox*{7cm}{!}{\includegraphics{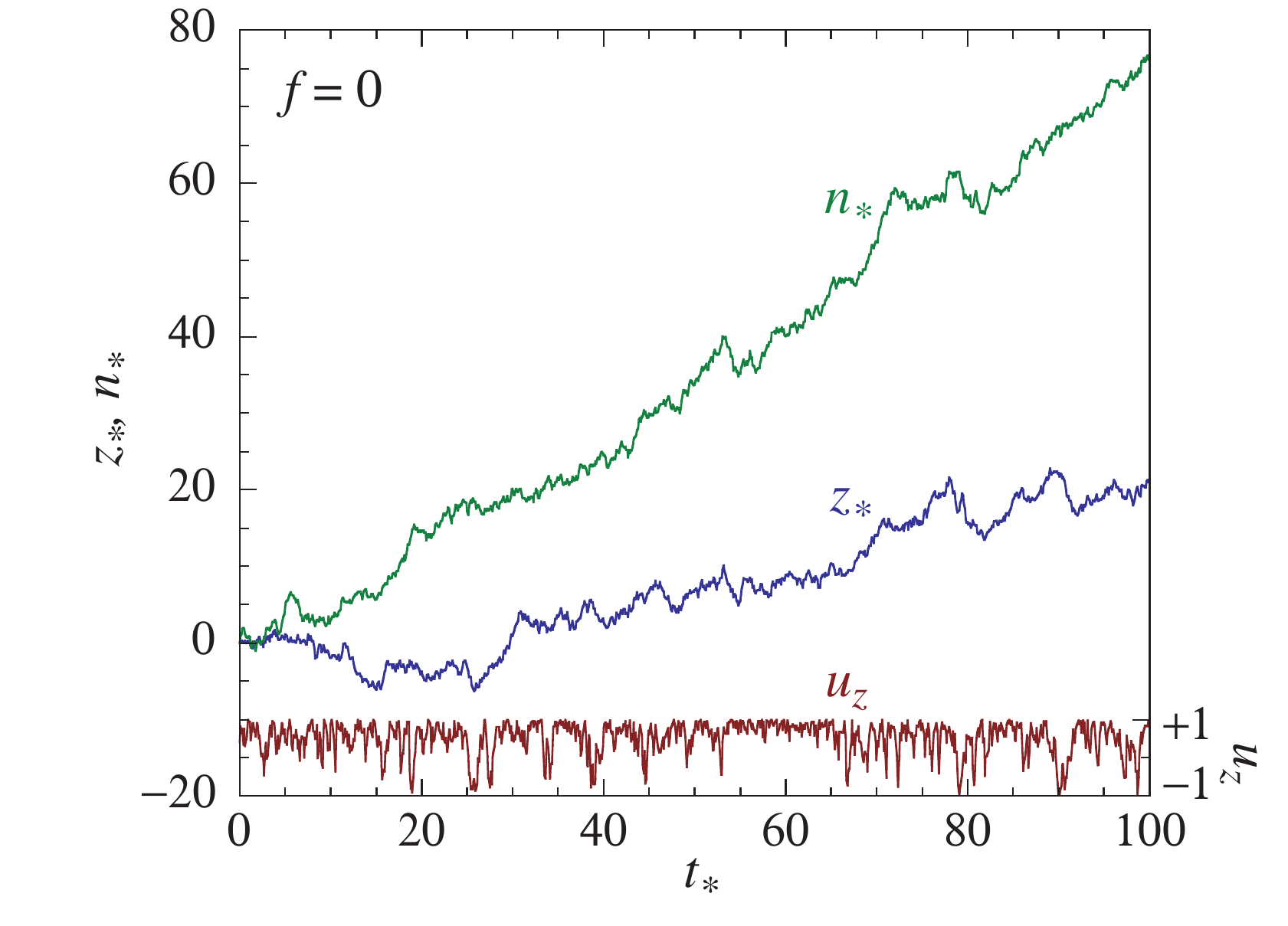}}
\resizebox*{7cm}{!}{\includegraphics{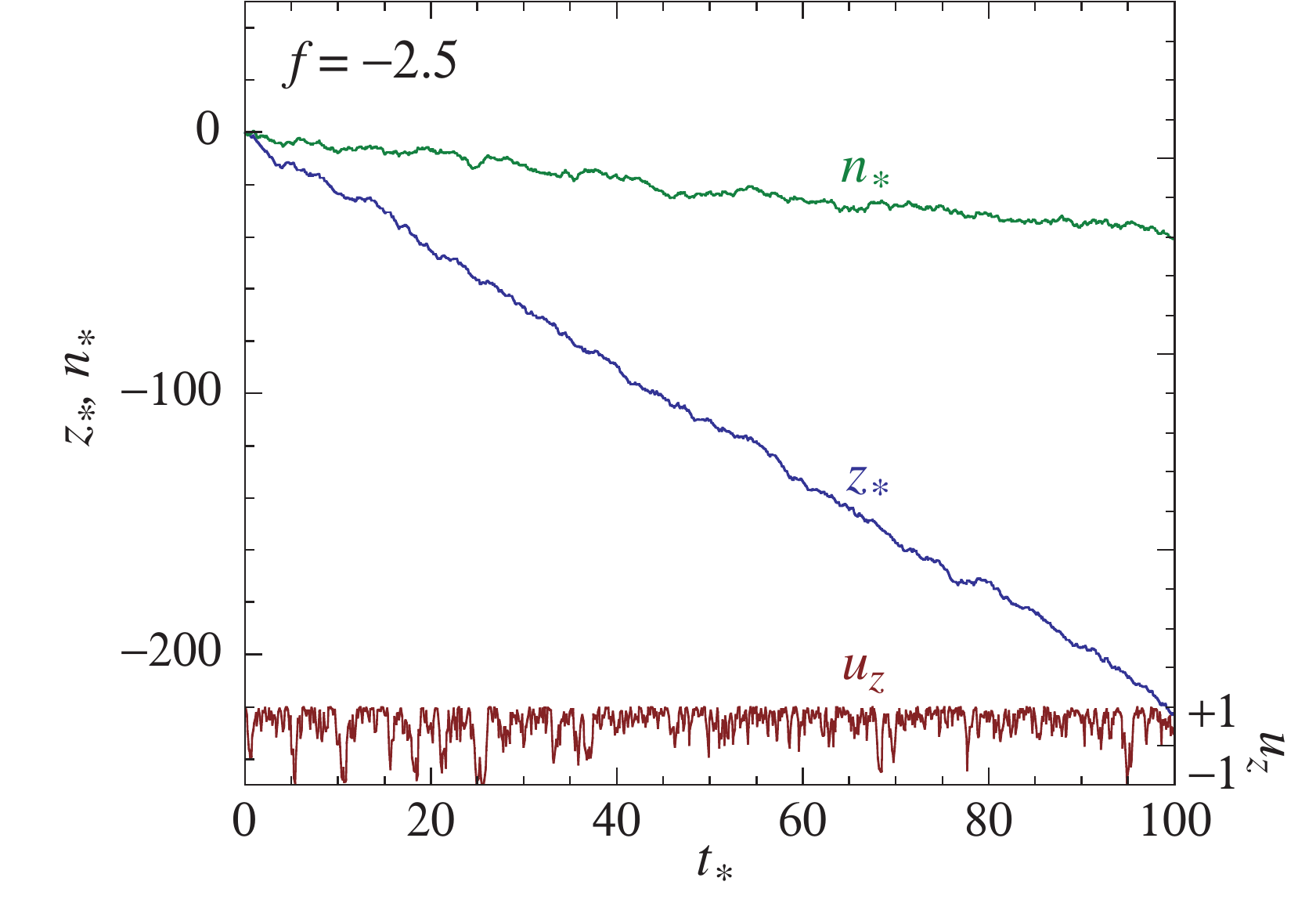}}
\caption{Janus particle subjected to an external force and magnetic field oriented in the $z$-direction: The random rescaled displacement $z_*=z/\sqrt{D_{\rm t}D_{\rm r}}$, the number of reactive events $\dot n_*=\dot n/\sqrt{D_{\rm rxn}D_{\rm r}}$, and the particle orientation $u_z$, versus the rescaled time $t_*=D_{\rm r}t$ for parameter values $\beta\mu B=2$, $W_{\rm rxn}/\sqrt{D_{\rm rxn}D_{\rm r}}=0.8$, and $\chi\sqrt{D_{\rm rxn}/D_{\rm t}}=0.8$ and (left) a zero external force, and (right) a rescaled external force equal to $f=\beta F\sqrt{D_{\rm t}/D_{\rm r}}=-2.5$. } \label{fig3}
\end{figure}

Examples of random trajectories obtained from simulations of these coupled overdamped Langevin equations are shown in Fig.~\ref{fig3}. In Fig.~\ref{fig3} (left), we see the self-propelled motion of the Janus particle by consumption of fuel in the absence of an external force. The particle is oriented by an external magnetic field in the $z$-direction. In Fig.~\ref{fig3} (right), an external force is exerted that is opposed to the direction of self-propulsion and large enough in magnitude so that the reaction is reversed and reactant is synthesized from product, instead of being consumed.

\begin{figure}[h]
\centering
\resizebox*{6cm}{!}{\includegraphics{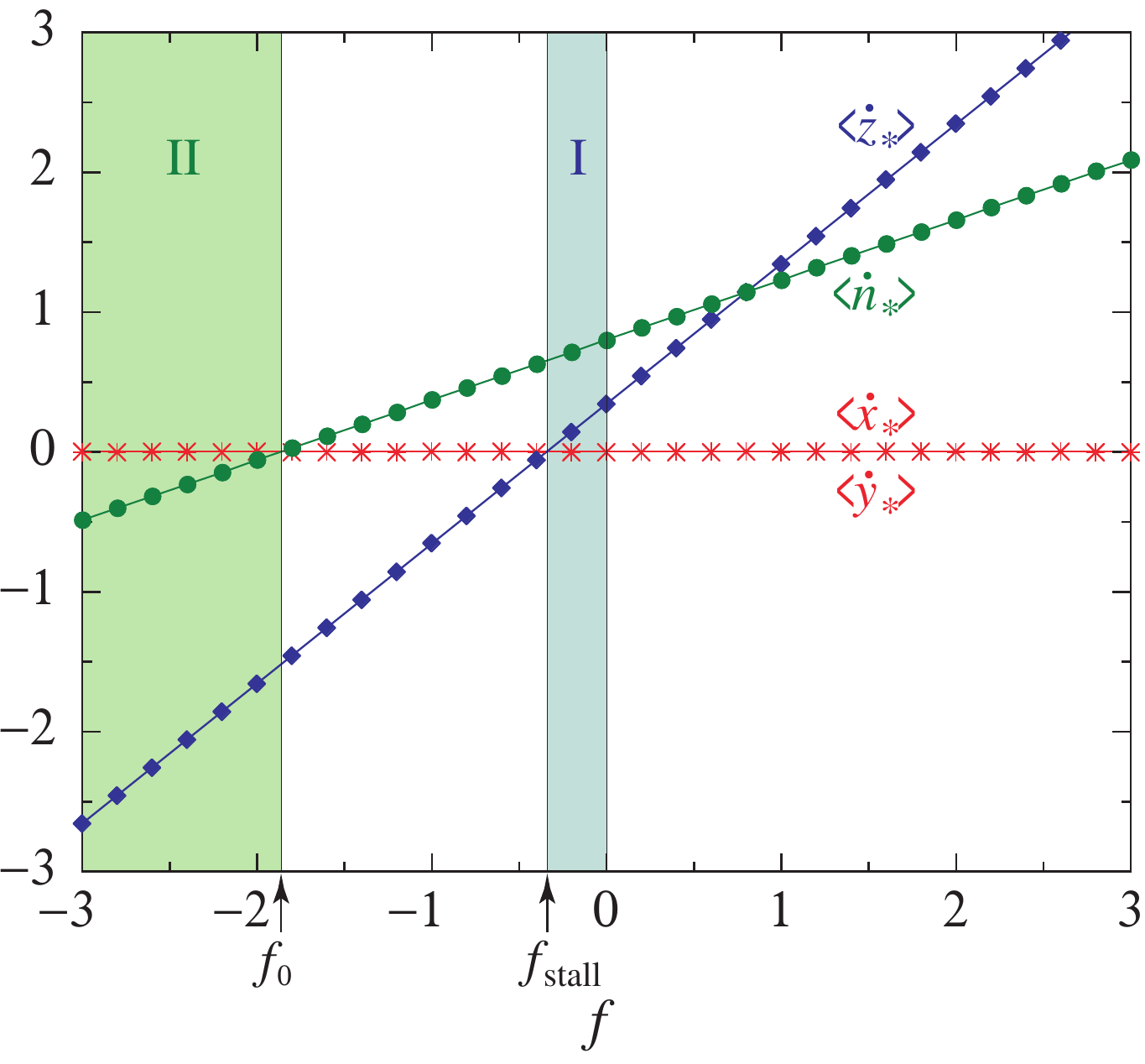}}
\caption{Janus particle subjected to an external force and magnetic field oriented in the $z$-direction \cite{GK17}: The mean values of the fluctuating rescaled velocities ${\bf\dot r}_*={\bf\dot r}/\sqrt{D_{\rm t}D_{\rm r}}$ and rate $\dot n_*=\dot n/\sqrt{D_{\rm rxn}D_{\rm r}}$ versus the rescaled magnitude of the external force $f=\beta F\sqrt{D_{\rm t}/D_{\rm r}}$ for the parameter values $\beta\mu B=2$, $W_{\rm rxn}/\sqrt{D_{\rm rxn}D_{\rm r}}=0.8$, and $\chi\sqrt{D_{\rm rxn}/D_{\rm t}}=0.8$.  The dots show the results of a numerical simulation with a statistics of $10^5$ trajectories integrated over the time interval $t_*=10$.  $f_{\rm stall}$ denotes the rescaled stall force and $f_0$ the threshold between fuel synthesis and consumption.  The Janus particle is propelled against the external force in the interval I.  Fuel synthesis happens in the interval II.} \label{fig4}
\end{figure}

If the external force and the magnetic field are oriented in the $z$-direction, ${\bf F}_{\rm ext}=(0,0,F)$ and ${\bf B}=(0,0,B)$, the particle is oriented on average in the same direction: $\langle u_z\rangle={\rm coth}(\beta\mu B)-1/(\beta\mu B)$, while the averages $\langle x\rangle=\langle y\rangle=0$ and $\langle u_x\rangle=\langle u_y\rangle=0$. Depending on the values of the mechanical and chemical affinities, the mean velocity $\langle\dot z\rangle$ and rate $\langle\dot n\rangle$ can take positive, vanishing, or negative values.  By definition the mean velocity vanishes at the stall force $F_{\rm stall}=-F_{\rm sd}\langle u_z\rangle$, which is proportional to the self-diffusiophoretic force $F_{\rm sd}=\gamma_{\rm t}\chi W_{\rm rxn}$. Also, the mean reaction rate is equal to zero at the force $F_0=-W_{\rm rxn}/(\beta\chi D_{\rm rxn}\langle u_z\rangle)$.  These two conditions are depicted in Fig.~\ref{fig4} that shows the mean values of the velocities and rate as a function of the rescaled external force $f$ for a positive value of the reaction rate $W_{\rm rxn}$.  In this case, the propulsion driven by the reaction exerts a mechanical work if the force is in the range $F_{\rm stall} <F<0$, corresponding to the domain I in Fig.~\ref{fig4}.  If the force is sufficiently opposed to propulsion to satisfy $F<F_0$, the mean reaction rate can become negative in the domain II of fuel synthesis from product.

Also, these theoretical predictions were compared with particle-based simulations of Janus particle dynamics~\cite{HSGK18}. The Janus particle is constructed from catalytic and non-catalytic beads that interact with the fluid species by repulsive intermolecular potentials while the motions of the fluid particles were described by multiparticle collision dynamics~\cite{MK99,K08,GIKW09}. The reactive and nonreactive dynamics of the system satisfied microscopic reversibility, and the nonequilibrium conditions were established by reservoirs with constant concentrations $\bar{c}_k$ far from the particle. The specification of the intermolecular potential and multiparticle collision parameters determined the transport properties of the system. As in the Langevin simulations, an external force along $z$ was applied and the magnetic field in the $z$-direction controlled the orientation of the Janus particle.

\begin{figure}[h]
\centering
\resizebox*{10cm}{!}{\includegraphics{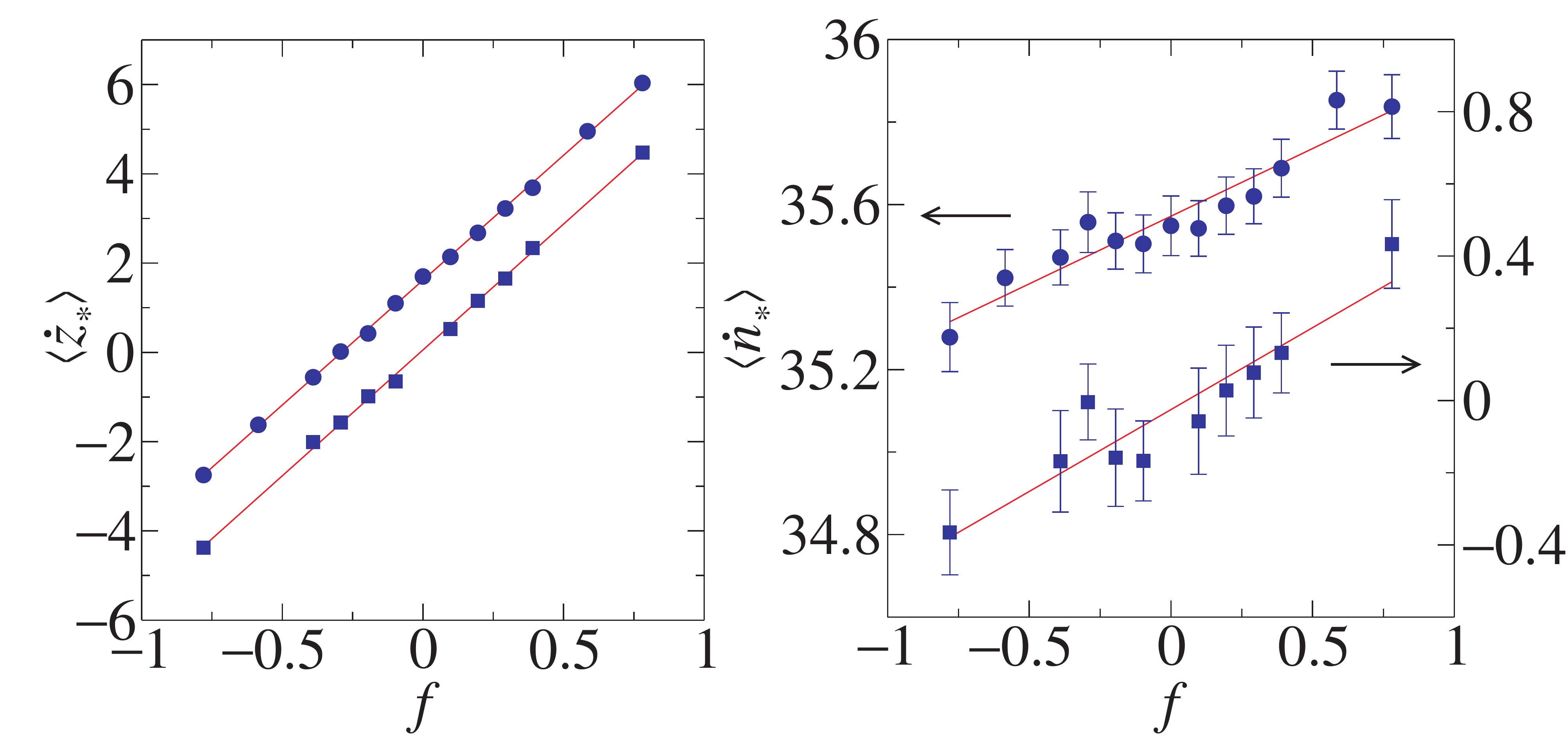}}
\caption{Simulations with a microscopically reversible kinetics of the Janus motor subjected to an external force $F_{\rm ext}$ and a magnetic field both oriented in the $z$-direction: Plots of the dependence on the rescaled force $f$ of the rescaled average motor velocity in the $z$-direction, $\langle \dot{z}_{\rm *} \rangle$ (left), and of the rescaled reaction rate, $\langle \dot{n}_* \rangle$ (right) for systems with (a) $\bar{c}_{\rm A} = 10$ and $\bar{c}_{\rm B} = 9$ (circles) and (b) $\bar{c}_{\rm A} = 10$ and $\bar{c}_{\rm B} = 10$ (squares). The results for (a) and (b) systems were obtained from averages over 200, 100 realizations of the dynamics, respectively. The fits to the data are indicated by (a) upper and (b) lower lines.  See \cite{HSGK18} for more information.} \label{fig5}
\end{figure}

The results of these simulations for $\langle \dot{z}\rangle$ and $\langle \dot{n}\rangle$ are shown in Fig.~\ref{fig5} as a function of the external force. The fits to these graphs indicated by the lines are in accord with the theoretical predictions. In particular we note that the reaction rate varies with the external force as predicted by the theory, $\langle \dot{n}\rangle= W_{\rm rxn} + \beta\chi D_{\rm rxn}\langle u_z\rangle F_{\rm ext}$. For the given parameters, $\beta\chi D_{\rm rxn}\langle u_z\rangle =0.006$ while the fit yields $0.006\pm 0.0006$. The results of another simulation, with parameters chosen so that the system is at equilibrium in the absence of an external force, are also presented in Fig.~\ref{fig5}. Since the diffusiophoretic parameter $\chi$ is nonzero, an external force can change the reaction rate. This is seen in the figure and we also see that for negative values of the external force the Janus particle converts product to fuel as predicted by the theory.

We note that nonlinear dependencies due to corrections in powers of the P\'eclet number manifest themselves for larger values of the external force magnitude.

\subsection{Thermodynamic efficiency}

Similar to molecular motors~\cite{JAP97}, the efficiency of the mechanical power of the Janus motor can be characterized by $\eta_{\rm m}\equiv-{\bf A}_{\rm mech}\cdot\langle{\bf\dot r}\rangle/(A_{\rm rxn}\langle \dot n\rangle)$,
and the efficiency of the chemical process of synthesis by $\eta_{\rm c}\equiv 1/\eta_{\rm m}$.
Since the thermodynamic entropy production rate of the coupled processes,
\be
\frac{1}{k_{\rm B}}\frac{d_{\rm i}S}{dt} ={\bf A}_{\rm mech}\cdot\langle{\bf\dot r}\rangle + A_{\rm rxn}\langle\dot n\rangle \geq 0\; ,
\label{entr-prod}
\ee
is non-negative according to the second law of thermodynamics, the mechanical and chemical efficiencies are bounded by $0\leq\eta_{\rm m}\leq 1$ and $0\leq\eta_{\rm c}\leq 1$ in their respective domains of application.

\begin{figure}[h]
\centering
\resizebox*{8cm}{!}{\includegraphics{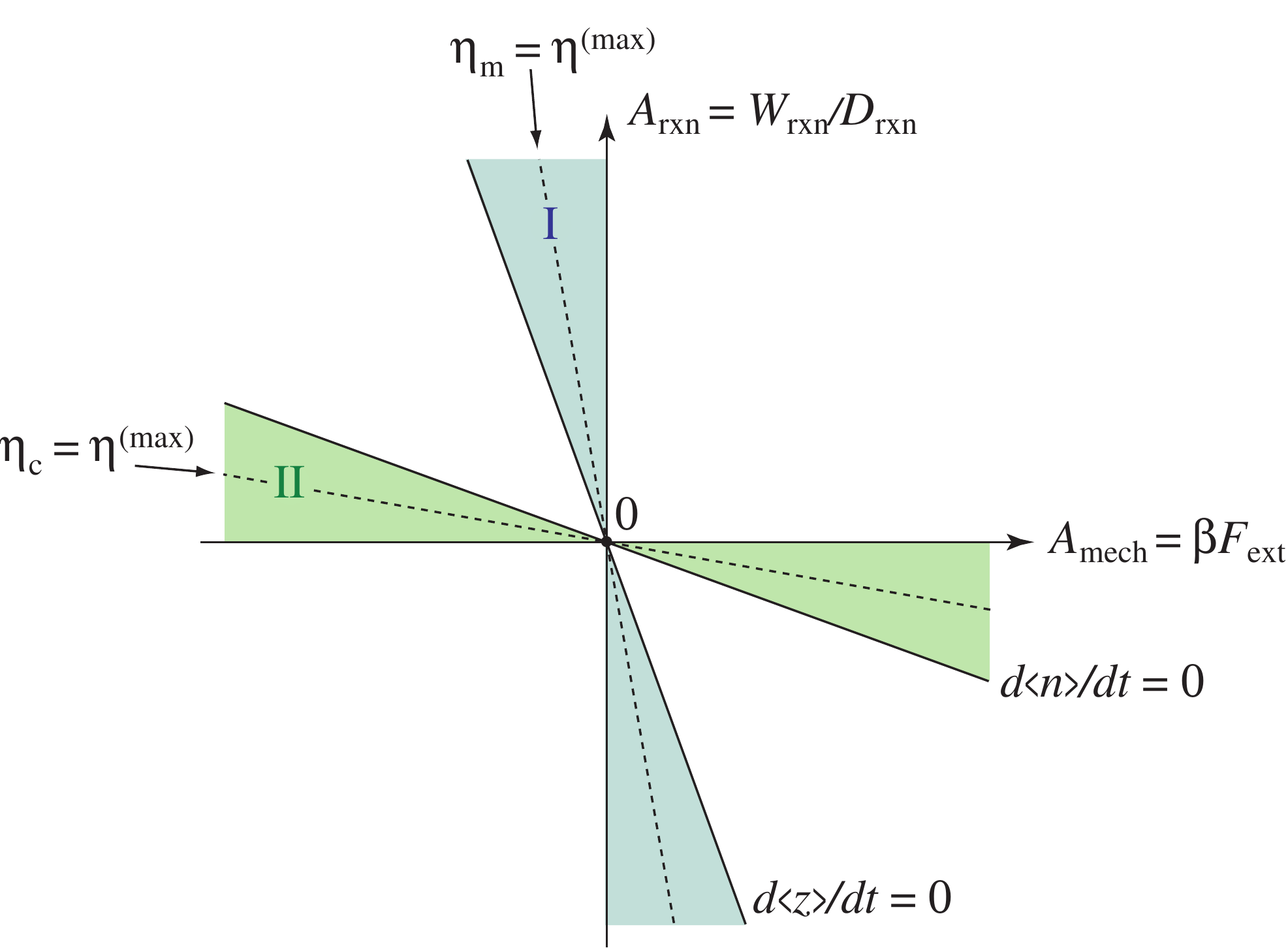}}
\caption{Schematic representation of the different regimes of the active particle in the plane of the mechanical and chemical affinities for $\chi>0$ \cite{GK18a}. In domain I, self-diffusiophoretic mechanical work is powered by the reaction.  In domain II, an external force of sufficient magnitude acting in a direction opposite to that of the Janus particle velocity can yield the synthesis of fuel from product.  For $\chi<0$, the slopes of the lines $\langle\dot z\rangle=0$ and $\langle\dot n\rangle=0$ are instead positive.} \label{fig6}
\end{figure}

The stall force $F_{\rm stall}$ where mean motor velocity vanishes and the force $F_0$ where the mean reaction rate vanishes are depicted in the ${\bf A}_{\rm mech}$-$A_{\rm rxn}$ plane of the mechanical and chemical affinities in Fig.~\ref{fig6}. Domain I in this figure corresponds to affinity values where self-diffusiophoretic propulsion occurs as a result of fuel consumption, while in domain II fuel is synthesised from product.

For a given chemical affinity, the maximal value of the mechanical efficiency is given by $\eta^{\rm (max)}\simeq \chi^2\langle u_z\rangle^2 D_{\rm rxn}/(4D_{\rm t})$.  Accordingly, the efficiency of self-propulsion increases as the square of the diffusiophoretic coupling $\chi$.  A similar expression holds for the chemical efficiency \cite{GK17}.  The locations where the efficiencies reach their maximal values are depicted as dashed lines in Fig.~\ref{fig6}.  The efficiency of self-propelled motors has been studied in~\cite{TK09} and in~\cite{SS12} by evaluating in detail the dissipation due to viscous friction in the surrounding solution.

\subsection{The Fokker-Planck equation and its consequences}

The Fokker-Planck equation that follows from the coupled overdamped Langevin equations~(\ref{eq-rot}) and (\ref{CL}) and governs the time evolution of the joint probability density $p({\bf X},{\bf u},t)$ can be written as
\bea
&&\partial_tp + \left(\chi \, W_{\rm rxn} \, {\bf u} + \beta D_{\rm t}\, {\bf F}_{\rm ext}\right)\cdot\pmb{\nabla} p + \left( W_{\rm rxn} + \beta\chi D_{\rm rxn} {\bf u}\cdot{\bf F}_{\rm ext}\right) \partial_np \nonumber\\
&&\qquad\qquad\qquad = D_{\rm t}\nabla^2 p + 2  \chi \, D_{\rm rxn} \, {\bf u}\cdot\pmb{\nabla}\partial_np + D_{\rm rxn}\partial_n^2p + \hat{L}_{\rm r} p\, ,
\label{FP-eq}
\eea
where the rotational diffusion operator is
\be
\hat L_{\rm r} p = D_{\rm r} \left\{\frac{1}{\sin\theta}\, \partial_{\theta} \left[ \sin\theta \, {\rm e}^{-\beta U_{\rm r}} \partial_{\theta}\left( {\rm e}^{\beta U_{\rm r}} p \right) \right] + \frac{1}{\sin^2\theta}\, \partial_{\varphi} \left[  {\rm e}^{-\beta U_{\rm r}} \partial_{\varphi}\left( {\rm e}^{\beta U_{\rm r}} p \right) \right] \right\},
\label{Lr}
\ee
with $U_{\rm r}= -\mu\, {\bf B}\cdot{\bf u}$ a rotational energy associated with the external torque exerted on the particle, for instance, due to an external magnetic field.

If we suppose that there is no external force ${\bf F}_{\rm ext}=0$ and integrate over the position~$\bf r$ and the orientation~$\bf u$ of the Janus particle, we obtain the equation, $\partial_t P + W_{\rm rxn} \partial_n P = D_{\rm rxn} \partial_n^2 P$, for the time evolution of the probability $P(n,t)$ that $n$ reactive events have happened during the time interval $[0,t]$. Therefore, the probability starting from $n=0$ at $t=0$ is Gaussian
\be
P(n,t) = \frac{1}{\sqrt{4\pi D_{\rm rxn} t}} \, \exp \left[ - \frac{(n-W_{\rm rxn}t)^2}{4 D_{\rm rxn} t}\right] ,
\label{n-Gaussian}
\ee
describing a random walk with a drift at the reaction rate $W_{\rm rxn}$ and the diffusivity~$D_{\rm rxn}$.

A similar result holds if the external force is non-vanishing.  Indeed, integrating over time the stochastic differential equation for $n$ in Eq.~(\ref{eq-n}) with an external force ${\bf F}_{\rm ext}=(0,0,F)$ in the $z$-direction and calculating the mean value and variance of $n(t)-n(0)$, we find the following effective rate and diffusivity,
\be
 W_{\rm rxn}^{\rm (eff)} = W_{\rm rxn} + \beta\chi D_{\rm rxn} F \langle u_z\rangle, \quad
D_{\rm rxn}^{\rm (eff)} = D_{\rm rxn} + (\beta\chi D_{\rm rxn} F)^2 t_{\Delta {\bf u}}  \, ,
\ee
in terms of the mean value of the orientation along the $z$-direction and the reorientation time $t_{\Delta {\bf u}}=\int_0^{\infty} dt \;\langle \Delta u_z(0)\Delta u_z(t)\rangle $, with $\Delta u_z=u_z-\langle u_z\rangle$.  By the central limit theorem, the probability $P(n,t)$ is given after a long enough time by the Gaussian~(\ref{n-Gaussian}) with the effective quantities instead of the bare ones.

Similarly, if we suppose that there is an external force ${\bf F}_{\rm ext}=(0,0,F)$ in the $z$-direction while the reaction rate vanishes $W_{\rm rxn}=0$ and integrate over the variables $x$, $y$, $n$, and $\bf u$, we find that the probability density to observe the particle with the position~$z$ at time~$t$ is also Gaussian
\be
{\mathscr P}(z,t) = \frac{1}{\sqrt{4\pi D_{\rm t} t}} \, \exp \left[ - \frac{(z-V_z t)^2}{4 D_{\rm t} t}\right] ,
\label{z-Gaussian}
\ee
here describing a random drift of mean velocity $V_z=\beta D_{\rm t}F$ and diffusivity $D_{\rm t}$.
Now, if the reaction rate is not equal to zero, we can integrate the stochastic differential equation~(\ref{eq-r}) over time and calculate the mean value and variance of $z(t)-z(0)$ to get the following effective drift velocity and diffusivity,
\be
V_{z}^{\rm (eff)} = \beta D_{\rm t}F + \chi W_{\rm rxn} \langle u_z\rangle,  \quad D_{\rm t}^{\rm (eff)} = D_{\rm t} + (\chi W_{\rm rxn})^2 t_{\Delta {\bf u}}, \label{Dt-eff}
\ee
also in terms of $t_{\Delta {\bf u}}$.
Again, by the central limit theorem, the probability density ${\mathscr P}(z,t)$ becomes after a long enough time the Gaussian~(\ref{z-Gaussian}) with the effective quantities instead of the bare ones.

In the absence of external magnetic field ${\bf B}=0$, there is no preferential orientation $\langle u_z\rangle=0$, so that the rotational motion remains diffusive and controlled by the rotational diffusion time
$t_{\rm rot}^{\rm (D)} = \int_0^\infty dt \, \langle {\bf u}(0)\cdot{\bf u}(t)\rangle_{\rm eq} = 3 \int_0^\infty dt \, \langle u_z(0)u_z(t)\rangle_{\rm eq} = 1/(2 D_{\rm r})$. In this case, we recover the known result \cite{K13} that the effective diffusion coefficient is given by
\be
D_{\rm t}^{\rm (eff)} = D_{\rm t} + \frac{V_{\rm sd}^2}{6 D_{\rm r}} = D_{\rm t} + \frac{\chi^2}{6 D_{\rm r}}\, W_{\rm rxn}^2 \, ,
\label{D_t_eff}
\ee
which is quadratic in the reaction rate $W_{\rm rxn}$ and thus also in the self-diffusiophoretic velocity~$V_{\rm sd}$.  The larger the magnitude of the self-diffusiophoretic velocity, the more enhanced the diffusive random walk of the Janus particle.

\section{Fluctuation theorems and microreversibility}
\label{sec:FT}

The fluctuation theorems are recent results of nonequilibrium statistical mechanics, showing that the ratio of the probabilities of opposite fluctuations in the particle or energy transfers $\Delta{\bf X}$ across an open system are related by
\be
\frac{P(\Delta{\bf X},t)}{P(-\Delta{\bf X},t)}\simeq_{t\to\infty} \exp({\bf A}\cdot\Delta{\bf X})
\ee
to the affinities or thermodynamic forces ${\bf A}$ driving the system out of equilibrium~\cite{LS99,AG04,AG07,J11,S12,G13NJP}.  These relations hold independently of the details of the process and arbitrarily far from thermodynamic equilibrium beyond the regimes of linear response.  At equilibrium, the affinities are equal to zero ${\bf A}=0$ and the ratio of the probabilities goes to one, so that we recover the principle of detailed balance.  The fluctuation theorems are established on the basis of the time-reversal symmetry of the underlying microscopic dynamics.  Remarkably, they imply that the entropy production
\be
\frac{1}{k_{\rm B}}\frac{d_{\rm i}S}{dt} =\lim_{t\to\infty} \frac{1}{t}\, \langle\Delta{\bf X}\rangle_{t} \cdot {\bf A} = \lim_{t\to\infty} \frac{1}{t} \int d\Delta{\bf X}  \, P(\Delta{\bf X},t) \, \ln \frac{P(\Delta{\bf X},t)}{P(-\Delta{\bf X},t)} \ge 0
\ee
is always non-negative in agreement with the second law of thermodynamics.

\subsection{Chemical fluctuation theorem}

A fluctuation theorem can be obtained for the Gaussian probability distribution~(\ref{n-Gaussian}).  Indeed, we have that
\be
\frac{P(n,t)}{P(-n,t)} = \exp\left( A_{\rm rxn} n\right)
\ee
with the chemical affinity $A_{\rm rxn} = W_{\rm rxn}/D_{\rm rxn}$, in the case without an external force when the mechanical affinity is equal to zero $A_{\rm mech}=0$.

\begin{figure}[h]
\centering
\resizebox*{6cm}{!}{\includegraphics{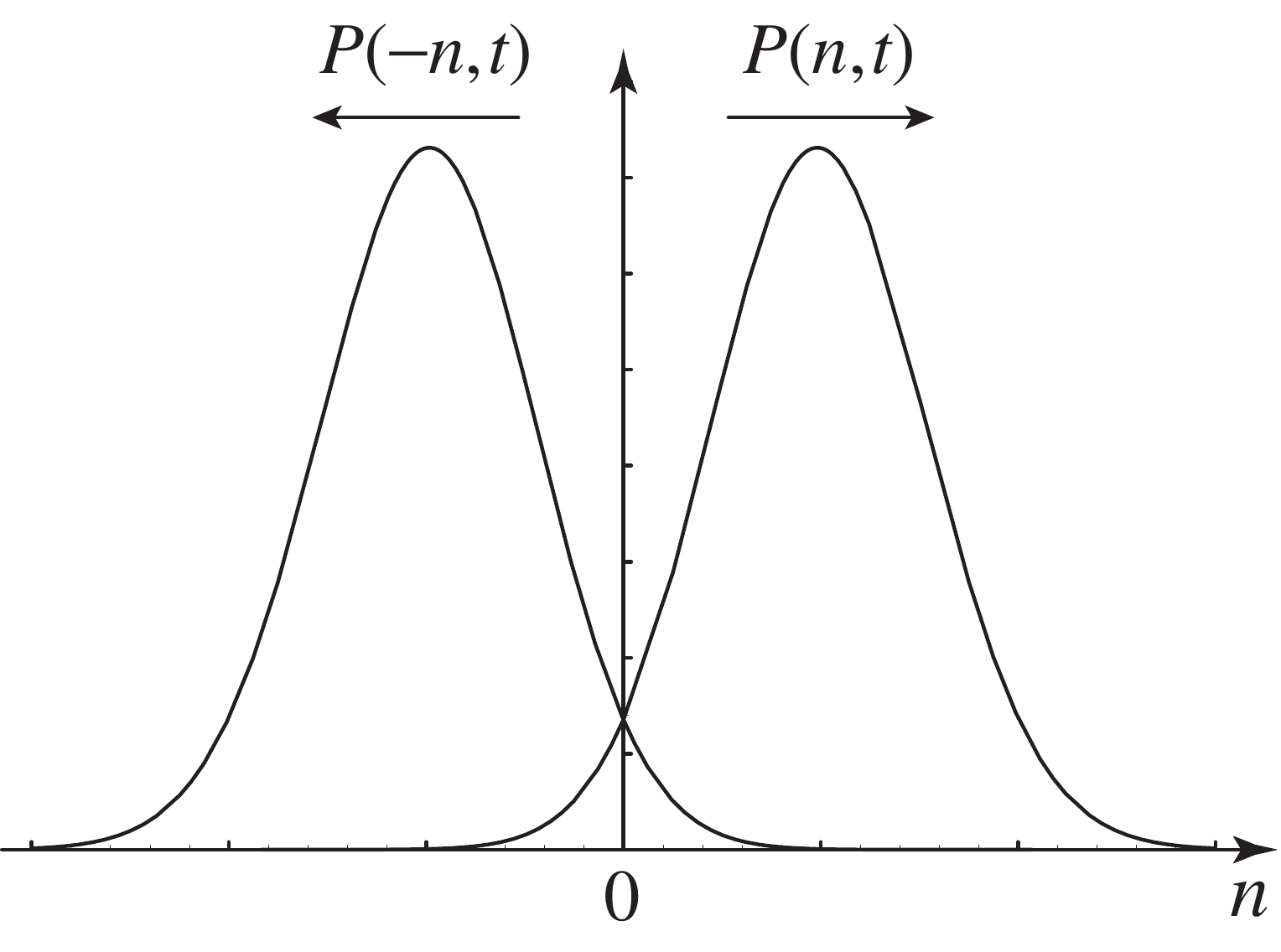}}
\caption{Schematic representation of the probability distributions $P(\pm n,t)$ of opposite fluctuations in the number $n$ of reactive events occurring during the time interval $[0,t]$ under nonequilibrium conditions.} \label{fig7}
\end{figure}

As shown in Fig.~\ref{fig7}, the probabilities $P(n,t)$ and $P(-n,t)$ to find the number $\pm n$ of reactive events at time $t$ shift away from the origin $n=0$  in opposite directions with the rates $\pm W_{\rm rxn}$.  Consequently, their overlap rapidly decreases as time $t$ increases. However, their ratio continues to vary as $\exp(A_{\rm rxn} n)$ in the long-time limit.  This remarkable property explains that the probability of observing an opposite fluctuation quickly becomes very small, so that the process is observed to be irreversible, although the opposite fluctuations are still present.

\subsection{Mechanical fluctuation theorem}

Using the Gaussian probability distribution~(\ref{z-Gaussian}), we have a similar fluctuation relation for the mechanical displacement $z$,
\be
\frac{{\mathscr P}(z,t)}{{\mathscr P}(-z,t)} = \exp\left( A_{\rm mech} z\right)
\ee
with the mechanical affinity $A_{\rm mech}=F/(k_{\rm B}T)$, in the case where $A_{\rm rxn}=0$.

By using the central limit theorem with the effective quantities~(\ref{Dt-eff}), we recover the effective fluctuation relation \cite{FPBCK16} for the displacement along the $z$-direction
\be
\frac{{\mathscr P}(z,t)}{{\mathscr P}(-z,t)} \simeq_{t\to\infty} \exp\left(\frac{F_{\rm eff}z}{k_{\rm B}T_{\rm eff}}\right) \, ,
\ee
which is expressed in terms of an effective force $F_{\rm eff}=F+F_{\rm sd}\langle u_z\rangle$ resulting from the external and self-diffusiophoretic forces, and the effective temperature $T_{\rm eff}=T\left[1+(V_{\rm sd}^2/D)t_{\Delta {\bf u}}\right]$.  Here, we see that fluctuation theorems can be used to justify the introduction of effective temperatures on the basis of nonequilibrium driving mechanisms.

\subsection{Mechanochemical fluctuation theorem}

\begin{figure}
\centering
\resizebox*{14cm}{!}{\includegraphics{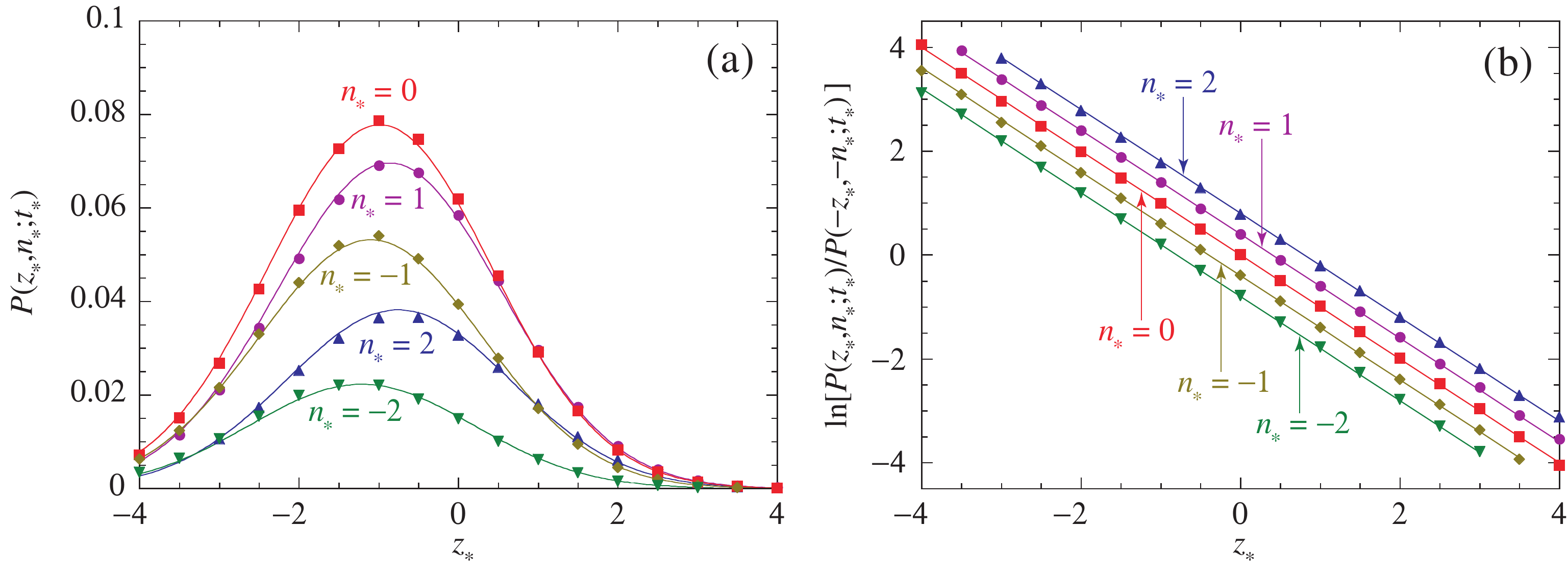}}
\caption{Janus particle subjected to an external force and magnetic field oriented in the $z$-direction \cite{GK17}: (a) Probability density $P(z_*,n_*;t_*)$ with $n_*=n\sqrt{D_{\rm r}/D_{\rm rxn}}=2,1,0,-1,-2$ versus the rescaled displacement $z_*=z\sqrt{D_{\rm r}/D_{\rm t}}$ at the rescaled time $t_*=D_{\rm r}t=1$ for the parameter values $p=\beta\mu B=1$, $f=\beta F\sqrt{D_{\rm t}/D_{\rm r}}=-1$, $w=W_{\rm rxn}/\sqrt{D_{\rm rxn}D_{\rm r}}=0.4$, and $c=\chi\sqrt{D_{\rm rxn}/D_{\rm t}}=0.4$. (b) Verification of the mechanochemical fluctuation relation~(\ref{FT}) in the same conditions.  The probability ratio is calculated if $P(z_*,n_*;t_*)$ and $P(-z_*,-n_*;t_*)$ are larger than $10^{-4}$. The dots are the results of a numerical simulation with an ensemble of $10^{7}$ trajectories and an integration with the time step $dt_*=10^{-3}$.  The lines depict the theoretical expectations.} \label{fig8}
\end{figure}

The previous results can be generalized to the joint probability ${\cal P}({\bf r},n,t) = \int d^2u \, p({\bf r},n,{\bf u},t)$ of a displacement $\bf r$ of the motor after $n$ reactive events have occurred during the time interval $[0,t]$, which obeys the mechanochemical fluctuation theorem
\be
\frac{{\cal P}({\bf r},n,t)}{{\cal P}(-{\bf r},-n,t)} \simeq_{t\to\infty} \exp\left({\bf A}_{\rm mech}\cdot{\bf r} + A_{\rm rxn} \, n \right) .
\label{FT}
\ee
This result has been proved in~\cite{GK17} using the method of the cumulant generating function by modifying the Fokker-Planck equation to include the counting variables.  The theorem~(\ref{FT}) implies the non-negativity of the entropy production~(\ref{entr-prod}).

Figure~\ref{fig8} shows that the mechanochemical fluctuation theorem is satisfied.  This theorem is also valid in the nonlinear regime away from equilibrium for reactions more complicated than those considered here, where it predicts generalizations of the Onsager reciprocal relations to the higher cumulants and their responses \cite{AG04,G13NJP}.

\subsection{Finite-time fluctuation theorem}

\begin{figure}
\centering
\resizebox*{6cm}{!}{\includegraphics{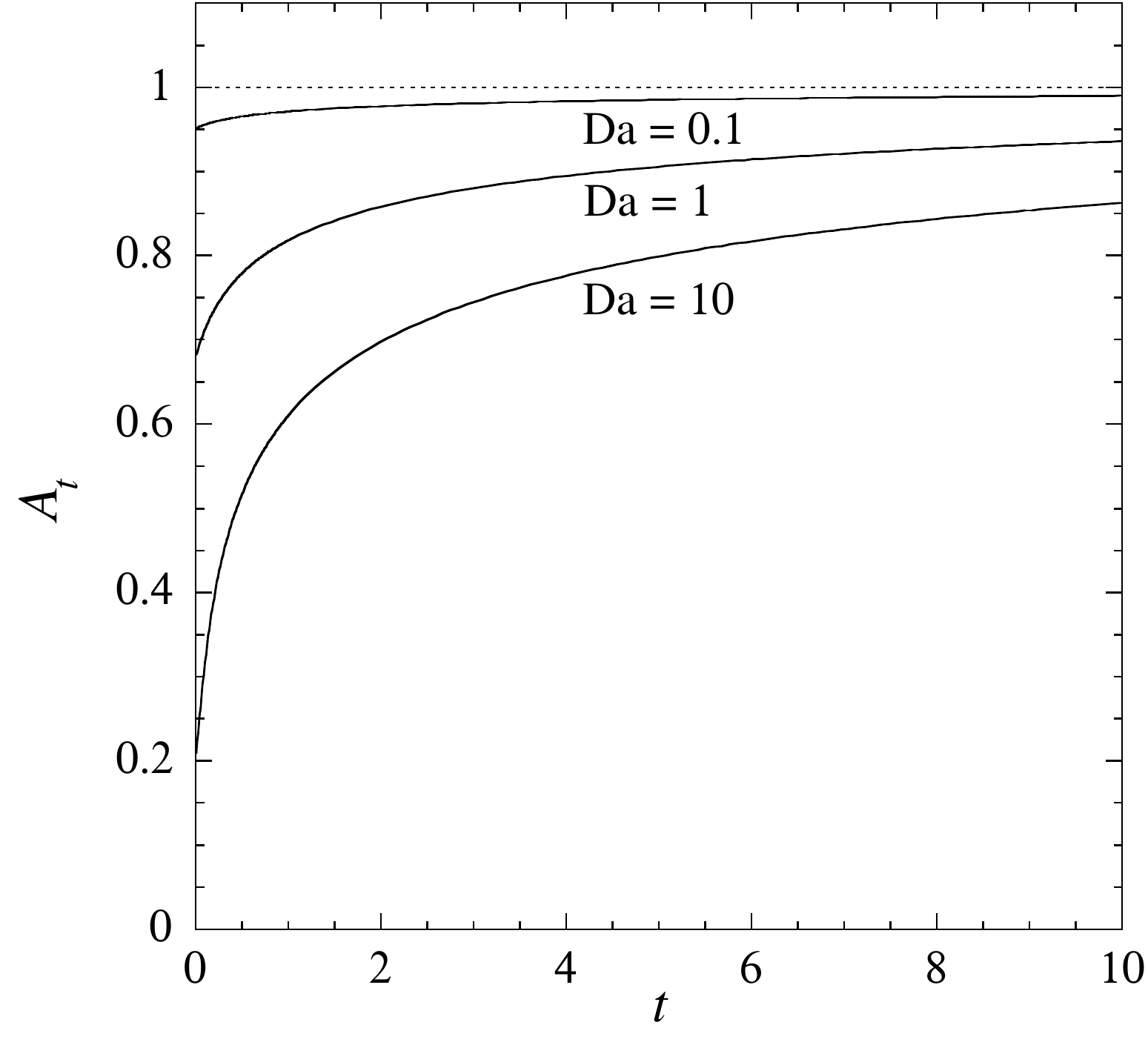}}
\caption{Time-dependent affinity $A_t$ of the finite-time fluctuation theorem for an immobile Janus particle of radius $R=1$ with surface reaction influenced by the diffusion of fuel and product from a spherical reservoirs at the distance $L=10$ from the particle center.  The asymptotic value of the affinity is equal to $A_{\rm rxn}=1$ (dotted line).  The diffusion coefficients take the value $D=D_{\rm A}=D_{\rm B}=1$ and the rate constants $\kappa=\kappa_+=\kappa_-$. The Damk\"ohler number is given by~${\rm Da}=2\kappa R/D$.} \label{fig9}
\end{figure}

In general, the aforementioned fluctuation theorems hold in the long-time limit.  They may also hold at finite times for the Gaussian processes of Eqs.~(\ref{n-Gaussian}) or~(\ref{z-Gaussian}).  Remarkably, this is also the case for more complicated processes involving linear reactions such as ${\rm A}\rightleftharpoons{\rm B}$ possibly coupled to diffusion~\cite{AG08,GK18c,GGHK18}.  Finite-time fluctuation theorems can be established for diffusion-influenced surface reactions on immobile catalysts.  These processes are ruled by the fluctuating diffusion equations~(\ref{diff-eq-1}) with the surface reaction rate~(\ref{ws}), but a fluid at rest (${\bf v}=0$).  In such systems, we have that $P(n,t)/P(-n,t) = \exp\left( A_t \, n\right)$ with a time-dependent affinity $A_t$ that converges towards $A_{\rm rxn}$ in the long-time limit.  Analytic expressions can be obtained for $A_t$ by solving a linear diffusion-reaction problem with special boundary conditions.

Figure~\ref{fig9} shows the time-dependent affinity for the diffusion-influenced surface reaction on an immobile Janus particle for three different values of the Damk\"ohler number.  We see that the convergence towards the asymptotic value is slower in the diffusion-limited than in the reaction-limited regime.

\section{Comparison with biomolecular motors and enzymes}
\label{sec:compare}

\subsection{Efficiencies and fluctuation theorems for biomolecular motors}

Energy transduction and mechanochemical coupling are general features common to colloidal and biomolecular motors~\cite{H05}.  Indeed, Fig.~\ref{fig6} that shows the efficiencies of a self-diffusiophoretic motor in the linear regime near equilibrium is comparable to similar diagrams obtained for molecular motors in the linear or nonlinear regimes \cite{JAP97,LLM08,A10, GG10}.  In these latter cases, the different domains in the plane of chemical and mechanical affinities are separated by curved instead of straight lines due to the kinetic nonlinearities.  Otherwise, for molecular motors, there also exists a domain where propulsion is powered by the reaction and another domain where fuel is synthesized by the external force, as in Fig.~\ref{fig6} for the self-diffusiophoretic motor.

Furthermore, mechanochemical fluctuation theorems hold for biomolecular motors~\cite{AG06b,LLM08,GG10}. Accordingly, the ratio of probabilities of opposite fluctuations in the movement and reaction of the motors behaves as
\be
\frac{P(x,n,t)}{P(-x,-n,t)} \simeq_{t\to\infty} \exp( A_{\rm mech}\, x + A_{\rm rxn}\, n )
\label{MCFT-motmol}
\ee
in terms of the mechanical and chemical affinities, $A_{\rm mech}$ and $A_{\rm rxn}$, which have similar definitions as for the self-diffusiophoretic motor.  These theorems are the expression of microreversibility and they have consequences for the Onsager reciprocal relations of the linear response coefficients, as well as their generalizations involving the nonlinear response coefficients \cite{AG04,AG07,G13NJP}.

These results show the analogy between self-diffusiophoretic active particles and molecular motors.  Similar results also hold for enzymes.

\subsection{Enhanced diffusion linear in the reaction rate}

According to Eq.~(\ref{D_t_eff}), the enhancement of diffusion by the self-diffusiophoretic mechanism is quadratic in the reaction rate.  We may wonder if there is a mechanism that is instead linear in the reaction rate.  Here, we consider propulsion by the release in a funnel of several product molecules from the fuel molecule.  As these molecules move away from the reaction center, the phase-space domain widens in the funnel as the translational, rotational, and vibrational degrees of freedom of the molecules are freed.  In this regard, the available phase-space volume increases as ${\cal V}(z)\sim z^f$ with the number $f$ of degrees of freedom and the distance $z$ from the reaction center in the funnel.  Therefore, the free-energy potential has an entropic contribution going as $G(z)\simeq -k_{\rm B}T\ln{\cal V}(z)$ and the force exerted on the motor is $F(z)=-\partial_z G(z)$ \cite{Z92,RR01,RLBK17}.  During the release, the motor is thus propelled according to $\gamma_{\rm t}dz/dt=F(z)$, giving the displacement $z=\sqrt{2f D_{\rm t}\Delta t}$ over the time interval $\Delta t$, where $D_{\rm t}=k_{\rm B}T/\gamma_{\rm t}$ is the diffusion coefficient corresponding to Stokes' friction coefficient $\gamma_{\rm t}$.  If the reaction proceeds at the rate $W_{\rm rxn}$, this displacement is repeated every time step $\Delta t\simeq 1/W_{\rm rxn}$ on average, generating the propulsion velocity $V\simeq \sqrt{2f D_{\rm t}W_{\rm rxn}}$.  Because of rotational diffusion at the rate $2D_{\rm r}$, the effective diffusion coefficient is thus given by
\be
D_{\rm t}^{\rm (eff)} =  D_{\rm t} + \frac{V^2}{6 D_{\rm r}} \simeq D_{\rm t} + \frac{4f}{9} \, R^2 \, W_{\rm rxn} \, ,
\ee
since the ratio of translational to rotational diffusion coefficients is equal to $D_{\rm t}/D_{\rm r}=4R^2/3$ in terms of   Stokes' hydrodynamic radius $R$ of the motor.  Therefore, this mechanism leads to a linear dependence on the reaction rate instead of the quadratic dependence of Eq.~(\ref{D_t_eff}).

\section{Collective motion}
\label{sec:collective}

Some of the most fascinating properties of self-propelled particles are associated with their collective behavior.  These include active self-assembly processes and a variety of dynamical cluster states, and there is a growing literature on this topic for systems of active particles~\cite{TCPYB12,EWG15,WDASM15,ZS16,BDLRVV16,LMC17}. We provide a sketch of how such collective behavior can be described but, in keeping with the theme of this review, we confine our attention to the collective behavior of systems containing colloidal motors that move by diffusiophoretic mechanisms.

Systems of self-propelled colloidal particles can experience different kinds of interactions: (1) Direct interparticle interactions already manifest themselves at equilibrium where they determine the phase diagram of colloidal systems \cite{J02}. They play no role in dilute suspensions, but they become increasingly important as the colloidal density increases. (2) Chemotactic interactions arise from nonequilibrium gradients of species concentration fields.  It is important distinguish between diffusiophoretic effects due to concentration gradients in the absence of reaction, and  self-diffusiophoretic effects in the presence of reaction. (3) Hydrodynamic interactions between particles are mediated by the velocity field.  They have a nonequilibrium origin since they involve the gradients of the velocity field; also hydrodynamic flows accompany the force-free dynamics of self-diffusiophoretic particles.

Here, we consider an isothermal dilute suspension of colloidal motors moving in a dilute solution of fuel A and product B molecular species.  Methods have been developed in order to perform an expansion in the concentration of colloidal motors \cite{E56,LL87,BKM77,LK79}.  At lowest order, the motors are isolated from each other and they move in a dilute solution of the molecular species.  As in the Chapman-Enskog expansion for the solution Boltzmann's equation, we may suppose that there exist local concentration gradients of fuel and product species and local velocity gradients of the fluid.  Consequently, we should first solve the problem where every colloidal particle moves in concentration and velocity fields that are not uniform far from the particle, but manifest concentration and velocity gradients.  To simplify the discussion, we assume that hydrodynamic effects play a negligible role and focus on the nature of the concentration fields.

\subsection{A colloidal particle moving in concentration gradients}

First, we have to solve the same problem as in Subsec.~\ref{sph-problem-RD}, but with the boundary condition $\pmb{\nabla}c_k\vert_{r=\infty} = {\bf g}_k$ at a large distance from the particle, instead of $c_k\vert_{r=\infty} = \bar{c}_k$.  The solution may again be expressed as in Eq.~(\ref{cA-f}), but with a different function $f$ now depending on the concentration gradients ${\bf g}_k$.  The force and the torque exerted on the Janus particle in such concentration fields can be calculated by Eqs.~(\ref{Fd-surf-av}) and~(\ref{rot-diff-t}).  In this way, we obtain the diffusiophoretic velocity and angular velocity,
\bea
&&{\bf V}_{\rm d} =\frac{{\bf F}_{\rm d}}{\gamma_{\rm t}}= \sum_k \zeta_k n_k \, {\bf u} + \sum_k \varepsilon_k \, \pmb{\nabla} n_k  +  \sum_k \theta_k \, {\bf u}\, {\bf u}\cdot \pmb{\nabla} n_k  \, , \label{Vd}\\
&&\pmb{\Omega}_{\rm d} =\frac{{\bf T}_{\rm d}}{\gamma_{\rm r}}= \sum_k \lambda_k {\bf u}\times\pmb{\nabla} n_k \, . \label{Wd}
\eea
The concentration gradients ${\bf g}_k$ have been written as $\pmb{\nabla} n_k$ because the concentration fields $n_k$ in the presence of many colloidal particles may differ on large scale from the local concentration profiles $c_k$ around every particle and, moreover, these gradients are no longer uniform on large scales as supposed with the notation ${\bf g}_k$, but are given instead by spatial derivatives of non-uniform fields $n_k$.

In order to have non-vanishing torque coefficients, $\lambda_k\neq 0$, we suppose that the diffusiophoretic constants $b_k$ are not uniform on the surface of the particle, so that Eq.~(\ref{rot-diff-t}) now gives the non-zero angular velocity~(\ref{Wd}). Moreover, this non-uniformity also contributes to the other coefficients $\zeta_k$, $\varepsilon_k$, and $\theta_k$.

If the diffusiophoretic constants $b_k$ are uniform on the particle surface, we have $\lambda_k=0$ and no angular velocity, $\pmb{\Omega}_{\rm d}=0$. Then, in a uniform background of concentrations, we have $\pmb{\nabla} n_k=0$ and the velocity reduces to the active self-diffusiophoretic velocity ${\bf V}_{\rm sd}$.  Instead, at chemical equilibrium where $\kappa_+ n_{\rm A}=\kappa_- n_{\rm B}$, the self-diffusiophoretic contributions to the force vanish, $\zeta_k=0$ and $\theta_k=0$, and there remains the passive diffusiophoretic velocity, ${\bf V}_{\rm d} =\sum_k \varepsilon_k \, \pmb{\nabla} n_k$ with $\varepsilon_k=b_k/(1+2b/R)$.  Out of chemical equilibrium,  the coefficients $\zeta_k$ are non-vanishing and there are active contributions of the reaction to the other coefficients $\varepsilon_k$ and $\theta_k$ as well.

\subsection{Ensemble of colloidal motors}

A dilute suspension of colloidal motors moving in a dilute solution of fuel A and product B molecular species may be described with the distribution function of the colloidal motors, $f_{\rm C}({\bf r},{\bf u}) \equiv \sum_{i=1}^{N_{\rm C}} \delta^3({\bf r}-{\bf r}_{i}) \, \delta^2({\bf u}-{\bf u}_i)$, expressed in terms of the positions and orientational unit vectors of the colloidal motors, $\{{\bf r}_i,{\bf u}_i\}_{i=1}^{N_{\rm C}}$. For a dilute suspension, the evolution equation of this distribution function can be deduced from the Fokker-Planck equation~(\ref{FP-eq}) for the probability that a single motor is located at the position $\bf r$ with the orientation $\bf u$ by using the diffusiophoretic velocity~(\ref{Vd}) and angular velocity~(\ref{Wd}) to get
\be
\partial_t f_{\rm C} +\pmb{\nabla}\cdot\left({\bf V}_{\rm d} \, f_{\rm C}-D_{\rm C}\pmb{\nabla}f_{\rm C}\right) = \hat L_{\rm r} f_{\rm C} \, ,
\label{master}
\ee
in terms of the effective diffusion coefficient $D_{\rm C}\simeq k_{\rm B}T/\gamma_{\rm t}$ and the operator~(\ref{Lr}) describing rotational diffusion of the colloidal motor in the effective rotational energy $U_{\rm r} = -\mu\, {\bf B}\cdot{\bf u} -\gamma_{\rm r}\sum_k \lambda_k \pmb{\nabla} n_k \cdot{\bf u}$, including the contribution of the diffusiophoretic torque in addition to the contribution from an external torque due to a magnetic field $\bf B$ acting on a magnetic dipole $\mu$ of the Janus particle.

Once, this equation is established, we can obtain the equations for the successive moments of $\bf u$, including the density of colloidal particles $n_{\rm C} \equiv \int d^2u \, f_{\rm C}({\bf r},{\bf u})$ and the polarizability or polar order parameter of the colloidal motors ${\bf p} \equiv \int d^2u \, {\bf u} \, f_{\rm C}({\bf r},{\bf u})$.  Higher moments can be neglected if only the smallest isotropies are considered.  These equations are coupled to the effective reaction-diffusion equations for the molecular species.

Such sets of coupled equations already manifest clustering instabilities, even without the coupling to hydrodynamics~\cite{SGR14,PS15}.  Collective dynamics of such diffusiophoretic motors has been studied in several types of system~\cite{TK12,CK17,HSK17,CRRK18}. The present methods provide a way to systematically deduce the equations of motion of these assemblies of colloidal motors, allowing their precise engineering.

\subsection{Coupling to hydrodynamics}

In order to study hydrodynamic interactions, we need to couple the previous diffusion-reaction equations to the Navier-Stokes equations for the velocity field and understand how these latter are modified by the presence of the colloidal particles.  It is already known that the viscosity coefficient is modified according to Einstein's formula and corrections~\cite{BKM77}, but we also need to determine how the activity of the motors affects the velocity field.  The methods developed since Einstein 1906 paper \cite{E56,LL87,BKM77} can be extended to determine not only the dissipative, but also the active contributions to the pressure tensor.  Several of them have already been obtained~\cite{R10}, but methods are available to deduce them systematically.

\section{Conclusion and perspectives}
\label{sec:conclude}

Thermodynamics and statistical mechanics provide fundamental approaches to study the propulsion mechanisms  of self-diffusiophoretic motors.  The coupling between motion and reaction required for energy transduction should be consistent with the underlying microreversibility of the processes considered.  This led to phenomena such as the prediction that the reaction rate for diffusiophoretic motors depends on the external force, and formed the basis for the fluctuation theorems for these systems. The phenomena described here should be observable in experiments.

The same general considerations enter in the description of collective behavior. The underlying many-motor dynamics must be consistent with microreversibility and, in addition, the manner in which the system is driven out of equilibrium will play a role in the forms that the collective behavior takes. The sketch of the formulation of this problem presented here will aid in the development of a full theory.

The study also shows that the design of colloidal motors requires engineering material interfacial properties, besides the shape of the colloids.  These properties, which are determined by the coating of the colloid, can be systematically identified with equilibrium and nonequilibrium interfacial thermodynamics.  They include the diffusiophoretic constants, together with the surface reaction rates, the slip length characterizing hydrophobicity, or the surface tension.

The approach can be extended to non-spherical particles, nonlinear surface reactions, electrodiffusiophoresis, or thermophoresis.  Among the other perspectives, one can envisage situations where the medium surrounding the active particle is a rarefied or dilute gas, instead of a liquid solution, as described by the Boltzmann equation, instead of the Navier-Stokes and diffusion equations.  The effects of long-time tails on diffusiophoresis could be investigated in such situations \cite{LR13}.  The study of these effects concerns transport and reaction of atmospheric aerosols \cite{LD95} or particles in interstellar clouds. The medium can also be a plasma, a viscoelastic fluid, or a liquid crystal, having different interfacial properties involved in the propulsion mechanism.

\section*{Acknowledgments}

The Authors thank Patrick Grosfils and Mu-Jie Huang for fruitful discussions. Financial support from the International Solvay Institutes for Physics and Chemistry, the Universit\'e libre de Bruxelles (ULB), the Fonds de la Recherche Scientifique~-~FNRS under the Grant PDR~T.0094.16 for the project ``SYMSTATPHYS", and the Natural Sciences and Engineering Research Council of Canada is acknowledged.


\end{document}